\documentclass[12pt]{article}
\usepackage{amssymb,amsmath}
\usepackage{epsfig}
\usepackage{color}
\usepackage{amsfonts}
\usepackage{graphics}
\usepackage{amscd}

\def\be{\begin{eqnarray}}
\def\ee{\end{eqnarray}}
\def\s{\sigma}

\def\half{\frac{1}{2}}

\def\a{\alpha}
\def\b{\beta}
\def\vev#1{\langle #1 \rangle}

\newcommand{\idn}{{1\relax{\kern-.35em}1}}

\newcommand{\hsp}[1]{\hspace{#1em}}

\DeclareMathOperator{\tr}{Tr} 

\begin{document}
\pagenumbering{arabic}
\begin{titlepage}
\begin{flushright}
{\tt hep-th/0602226}, WIS/03/06-FEB-DPP
\end{flushright}
\vspace{7 mm}
\begin{center}
 {\huge On the Worldsheet Theories of Strings Dual}
 \vspace{5 mm} {\huge to Free Large $N$ Gauge Theories}
\end{center}
\vspace{5 mm}
\begin{center}
{\large
Ofer Aharony\footnote{{\tt
Ofer.Aharony@weizmann.ac.il}.}$^\clubsuit$ , Zohar
Komargodski\footnote{{\tt Zkomargo@weizmann.ac.il}.}$^\clubsuit$
and Shlomo S. Razamat\footnote{{\tt
Razamat@physics.technion.ac.il}.}}$^\spadesuit$\\
 \vspace{3mm} $^\clubsuit$ Department of
Particle Physics, The Weizmann Institute of Science,\\ Rehovot
76100, Israel\\
$^\spadesuit$ Department of Physics, Technion -- Israel Institute
of Technology,\\ Haifa 32000, Israel
\end{center}
\vspace{2mm}
\noindent We analyze in detail some properties of the worldsheet
of the closed string theories suggested by Gopakumar to be dual to
free large $N$ $SU(N)$ gauge theories (with adjoint matter
fields). We use Gopakumar's prescription to translate the
computation of space-time correlation functions to worldsheet
correlation functions for several classes of Feynman diagrams, by
explicit computations of Strebel differentials. We compute the
worldsheet operator product expansion in several cases and find
that it is consistent with general worldsheet conformal field
theory expectations. A peculiar property of the construction is
that in several cases the resulting worldsheet correlation
functions are non-vanishing only on a sub-space of the moduli
space (say, for specific relations between vertex positions).
Another strange property we find is that for a conformally
invariant space-time theory, the mapping to the worldsheet does
not preserve the special conformal symmetries, so that the full
conformal group is not realized as a global symmetry on the
worldsheet (even though it is, by construction, a symmetry of all
integrated correlation functions).
\vspace{7mm}
\begin{flushleft}
February 2006
\end{flushleft}
\end{titlepage}
\tableofcontents
\section{Introduction and Summary}
It is widely believed that large $N$ $SU(N)$ gauge theories (with
adjoint matter fields) are dual to closed string theories with a
string coupling constant $g_s \sim 1/N$, in the limit of large $N$
with fixed 't Hooft coupling $\lambda \equiv g_{YM}^2N$. The
original argument of 't Hooft for this duality
\cite{'tHooft:1973jz} was based on a reinterpretation of the
Feynman diagrams of the gauge theory as closed string diagrams.
The Feynman diagrams may be written in 't Hooft's double-line
notation, in which they can be interpreted as two dimensional
surfaces with holes. It was conjectured that there should be an
equivalent description in which the holes get filled up, leading
to closed Riemann surfaces without boundaries.
In a normalization in which the gauge coupling constant appears
only as a factor of $1/g_{YM}^2$ sitting in front of the action,
the dependence of each Feynman diagram on $g_{YM}$ and on the rank
$N$ of the gauge group is determined by the topology of its
surface. A graph with $V$ vertices and $E$ propagators, whose
topology has $g$ handles and $h$ holes, is proportional to
\begin{equation}
(g_{YM}^{2})^{-V+E}N^h=(g_{YM}^2)^{-V+E-h} (g_{ YM}^2 N)^h=(g_{
YM}^2)^{2g-2} (g_{ YM}^2 N)^h=(g_{ YM}^2)^{2g-2}\lambda^h.
\end{equation}
For any correlation function $M$, the sum over Feynman diagrams
may then be rewritten as a sum over all topologies,
\begin{equation}\label{naive open closed}
    M=\sum_{g= 0}^\infty
\sum_{h=1}^\infty
F_{g,h}(g_{YM}^2)^{2g-2}\lambda^h=\sum_{g=0}^{\infty}
F_g(\lambda)(g_{YM}^2)^{2g-2}=\sum_{g=0}^{\infty} {\tilde
F}_g(\lambda) N^{2-2g},
\end{equation}
where $F_g(\lambda)\equiv \sum_{h=1}^\infty \lambda^h F_{g,h}$,
${\tilde F}_g(\lambda) \equiv \lambda^{2g-2} F_g(\lambda)$, and
the coefficients $F_{g,h}$ depend on all other coupling constants
of the theory. This equation has the same form as the perturbative
expansion of a closed string theory with string coupling $g_s \sim
1/N$, motivating the conjecture described above.
Even though the derivation of (\ref{naive open closed}) was based
on perturbation theory, such an expansion is believed to exist for
any value of $\lambda$.

The arguments above do not give a direct construction of the
string theory dual to a specific large $N$ gauge theory. In the
beginning, the only examples of 't Hooft's conjecture were limited
to field theories in two dimensions or less. This has changed in
the last decade, following \cite{Maldacena:1997re}. Now there are
many examples in which it is known how to find the closed string
dual of gauge theories which can be realized as the world-volume
theories of D-branes in some decoupling limit\footnote{There are
also more general dualities between open and closed string
theories, which we will not discuss here.}. In these
cases\footnote{This is believed to be true also for general gauge
theories, which can be reached by deformations of theories living
on D-branes.} the closed string dual turns out to be a standard
closed string theory, living in a warped higher dimensional space.
In some cases, for which the gauge theory is strongly coupled, the
dual string background is weakly curved and a gravity
approximation of the string theory may be used. In general (and,
in particular, for all weakly coupled gauge theories), this is not
the case, and the dual string theory is complicated (and does not
necessarily have a geometrical interpretation). For standard gauge
theories it is not known how to derive the duality to closed
strings, though there is a lot of evidence that it is correct; in
some topological cases one can provide an explicit derivation of
the duality \cite{GopakumarVafa}.
%
%

The general mapping (\ref{naive open closed}) works for any value
of $\lambda$, and in particular it is interesting to consider the
$\lambda \to 0$ limit, for which the gauge theory has a
perturbative expansion. This limit does not always make sense,
since in many cases $\lambda$ is related to the only scale in the
theory. However, the limit $\lambda \to 0$ is expected to be
smooth in many cases of four dimensional conformal gauge theories
(which are conformal for every value of the gauge coupling), such
as the ${\cal N}=4$ supersymmetric Yang-Mills theory (dual to type
IIB string theory on $AdS_5\times S^5$), and even in other
theories one can try to use (\ref{naive open closed}) to define a
closed string theory when $\lambda$ is strictly equal to
zero\footnote{Of course, the free gauge theory does not have
confining strings, which were one of the original motivations for
't Hooft's proposal. However, following the AdS/CFT
correspondence, it was recognized that confinement is not a
necessary feature for a dual string theory to exist. Compactified
free gauge theories do exhibit a deconfinement phase transition
and a Hagedorn spectrum \cite{ConfDeconf}.}. Indeed, the
correlation functions of free gauge theories (with $g_{YM}=0$)
have a topological expansion of the form (\ref{naive open closed})
in powers of $1/N^2$, and it is interesting to ask what is their
closed string dual\footnote{Of course, we are discussing here
correlation functions of local gauge-invariant operators such as
$\tr(F_{\mu \nu}^n)$, and not the S-matrix which is the identity
matrix in the free gauge theory. These are expected to be the
correct observables for gauge theories and for their string theory
duals.}. Note that in this limit the closed string coupling must
be identified with $1/N$ rather than with the vanishing $g_{YM}$
(even though D-brane constructions usually give $g_s \propto
g_{YM}^2$).  Clearly, given a dual to the free gauge theory, one
can map the interaction vertices in space-time to interactions on
the worldsheet (by the mapping of gauge-invariant operators in
space-time to integrated vertex operators on the worldsheet), and
rewrite the perturbative expansion in $\lambda$ of the space-time
gauge theory as a perturbative expansion in $\lambda$ on the
worldsheet.

There have been various proposals for how to study the string dual
of free large $N$ gauge theories (see, for instance,
\cite{FreeFields,StringBits,Joe,Karch,Bonelli:2004ve,
Itzhaki:2004te,Gopakumars,Bianchi,
Akhmedov}). It is clear that the dual string theories must live in
a highly-curved background, which may or may not have a
geometrical interpretation (for four dimensional free gauge
theories with massless adjoint fields, which are conformally
invariant, one expects that any geometrical interpretation should
include an $AdS_5$ factor). In this paper we will study in detail
a specific proposal by R. Gopakumar \cite{Gopakumars} for how to
map the Feynman diagrams to worldsheets\footnote{This proposal was
further studied in \cite{Furuuchi:2005qm}.}. This proposal is
based on rewriting the propagators in the Feynman diagrams as
integrals over Schwinger parameters, and mapping these parameters
to the moduli of a Riemann surface with holes (which include the
moduli of the closed Riemann surface, plus the circumferences of
the holes). One can then integrate over the parameters of the
holes, and translate any Feynman diagram to a correlation function
on the string worldsheet. We will focus on the special case of
correlation functions of gauge-invariant operators involving
adjoint scalar fields in four dimensional free gauge theories, but
the conclusions apply more generally, to any local gauge-invariant
operators in any free gauge theory.

The mapping of \cite{Gopakumars} (which we review below) gives a
closed string theory whose integrated correlation functions (of
physical vertex operators), by construction, reproduce the
space-time correlation functions. The worldsheet theory is also
automatically conformally invariant (so that it can be interpreted
as a closed string theory in conformal gauge) and modular
invariant. However, the construction does not give a Lagrangian
for the worldsheet theory, and it is not clear from the
construction if this worldsheet theory is a standard local
conformal field theory or not.

In this paper we will note two strange properties of the
worldsheet correlation functions resulting from Gopakumar's
prescription. The first is that, for some Feynman diagrams, the
correlation functions turn out to be non-vanishing only on a
sub-space of the moduli space. The dimension of the moduli space
for a closed string $n$-point function at genus $g$ is $2n+6g-6$,
and in some cases we find that the locus on which the correlation
function is non-vanishing has a lower dimension. For example, for
a particular planar 4-point function, we find that it is
non-vanishing only when all four points lie on a line. Obviously,
such a result is not consistent with the usual analyticity
properties of correlation functions in local field theories (such
as having operator product expansions at small distances). There
are (at least) three possible explanations of this result :
(a)~Gopakumar's prescription is wrong; (b)~The worldsheet theory
is non-local, and its correlation functions are not analytic;
(c)~There are global contributions to correlation functions (in
addition to the local ones), coming for instance from zero modes
on the worldsheet, which cause the correlation functions to vanish
in many circumstances. We will not be able to determine which of
these interpretations is correct -- this is an interesting topic
for further research.

A second strange property involves the space-time symmetries.
Apriori one would expect that any (global) symmetry of the
space-time theory should be realized as a symmetry of the
worldsheet theory, which does not act on the worldsheet
coordinates. For example, in the AdS/CFT correspondence, both
global symmetries and the conformal symmetry in space-time map to
global symmetries on the worldsheet (often related to isometries
of a sigma model). Gopakumar's prescription guarantees that any
space-time symmetries will be present in the integrated
correlation functions, but it does not guarantee that they will
act locally on the worldsheet. We will find that, for conformally
invariant theories, the mapping preserves the Poincar\'e and
scaling symmetries, but not the special conformal transformations.
These transformations are not manifestly preserved by the mapping,
and by explicit computations we show that they are not symmetries
of the worldsheet correlation functions (at fixed positions on the
worldsheet). Thus, it seems that (unlike in the AdS/CFT
correspondence) the Poincar\'e and scaling symmetries are realized
as global symmetries on the worldsheet, but the special conformal
symmetries are not. It would be interesting to investigate
possible modifications of Gopakumar's prescription in which the
special conformal transformations would also act locally on the
worldsheet (this may require a different choice of gauge for the
worldsheet diffeomorphisms than the one implied by the mapping of
\cite{Gopakumars}, just like Lorentz transformations do not act
locally in light-cone gauge and space-time supersymmetry does not
act locally in the NSR superstring formulation).

On the more positive side, we investigate in detail the operator
product expansion (OPE) on the worldsheet, resulting from specific
correlation functions in the limit where two points on the
worldsheet approach each other. As mentioned above, in some cases
this OPE is ill-defined because the correlation functions are
non-vanishing only at special positions, but in other cases the
correlation functions are non-zero when the two points approach
each other, so one can check if they have an OPE expansion. We
find that, in such cases, there is indeed a sensible OPE
expansion, as expected from a standard local worldsheet theory. We
compute the worldsheet dimensions of the operators appearing in
the OPE, we show that they are consistent with a local conformal
field theory (namely, they have integer values of $h-{\bar h}$),
and we show that consistent results emerge from different
correlation functions (of the same operators). This may be viewed
as support for option (c) above, that the conformal worldsheet
theory is a standard local theory, but with some global
contributions to its correlation functions.

We begin in section 2 by reviewing Gopakumar's mapping from
Feynman diagrams to worldsheets in detail. The mapping involves
Strebel differentials, and we review what these are and how they
are related to the moduli space of Riemann surfaces with holes. We
also discuss the strange properties noted above. Section 3 is a
general discussion of our expectations about space-time and
worldsheet OPEs for free gauge theories and their string duals. In
section 4 we discuss in detail several examples of planar
correlation functions, with three, four, and five vertices. We
show how to map the Feynman diagrams to worldsheet correlation
functions (by computing the relevant Strebel differentials) and
analyze their OPE expansions. In section 5 we perform a similar
analysis for the two-point function on the torus. In appendix A we
provide a few more examples of planar correlation functions, for
which we were not able to perform the full computation. Finally,
in appendix B we review some properties of elliptic functions
which are used in section 5.

\section{A Recipe for the Closed String Dual of Free Gauge Theories}
\label{Preliminaries}
\subsection{Mathematical background}
The nuts and bolts of Gopakumar's proposal rely on the mathematical theory
of Strebel differentials, and in particular on a theorem
by K. Strebel which we describe below\footnote{Strebel differentials are useful
also in closed string field theory \cite{Zwiebach,Moeller:2004yy}.}.
Let $X$ be a Riemann surface. A
quadratic differential is an expression of the form $q=\phi(z)dz^2$,
which is a section of the bundle
$T_X^{\mathbb{C}}\bigotimes T_X^{\mathbb{C}}$ (it may have poles
for marked surfaces). Given such a differential, we call a curve
$\gamma(t)$ horizontal if it satisfies
$\phi(\gamma(t))(\gamma ' (t))^2>0$.
Similarly, a vertical curve
satisfies $\phi(\gamma(t))(\gamma ' (t))^2<0$. The prime denotes
differentiation with respect to the affine parameter of the curve.
The horizontal curves have an interesting behavior near the zeros
and poles of $\phi$.
At a zero of order $m$ there are $m+2$ horizontal curves
intersecting. At a simple pole there are no locally intersecting
horizontal curves. Near a pole of second order, one can locally write
\begin{equation}\label{secondorderpole} \phi(z)dz^2 \simeq
-\frac{p^2}{(2\pi)^2z^2}dz^2.
\end{equation}
To classify the horizontal and vertical curves, we note that the circular
curves
\begin{equation}
\gamma(t)=r_0e^{it}, \hsp{1.5} t\in [0,2\pi)
\end{equation}
are horizontal near $z=0$, while
straight lines emanating from the pole,
\begin{equation}
\gamma(t)=te^{i\theta},\hsp{1.5} t\in \mathbb{R}_+
\end{equation}
are vertical near $z=0$. Thus,
locally around any pole of second order the geometry is that of a
semi-infinite cylinder.

From here on we will focus on the horizontal curves, whose global
structure may in general be very complicated.  It is customary to
distinguish the {\it closed} simple horizontal curves from the
other horizontal curves, and to call the complimentary set to the
closed simple horizontal curves the set of critical curves (in
particular, these are the curves that go through the intersection
points at the zeros of $\phi$). The set of critical curves for a
general second order differential is very complicated and not much
is known about it. Here the notion of a Strebel differential is
important. Let there be $n$ marked points $x_1,\cdots,x_n$ on our
compact Riemann surface $X$. A {\it Strebel differential} $q$ is a
second order meromorphic differential which satisfies the
following requirements :
\begin{itemize}
\item $q$ is holomorphic on $X\setminus \{x_1,\cdots,x_n\}$,
\item $q$ has a pole of second order at each $x_i$,
\item The set of critical curves of $q$ is compact and of measure zero.
\end{itemize}

Now we are ready to quote the theorem of K. Strebel (for proofs
and more details see
\cite{K.Strebel:1984,Mulase:98,Zvonkine:2002}). Assigning a
positive number $p_i$ to each marked point $x_i$, the theorem
states that there is a unique Strebel differential with double
poles of residues $p_i$ at the points $x_i$. Equivalently, we
require $\oint_{x_i} \sqrt{q}=p_i$, where the integration is on a
horizontal closed curve (close enough to the double pole) and the
branch cut is chosen so that the integral is positive with respect
to the orientation specified by the complex structure. The space
of possible $X$'s with $n$ marked points and a positive number
$p_i$ at each marked point is called the decorated moduli space of
Riemann surfaces, $\mathcal{M}_{g,n}\times \mathbb{R}_+^n$, where
$g$ is the genus of $X$. The theorem provides a unique Strebel
differential for any point on this moduli space.

The conclusions above can be carried further to establish a more
useful (for us) isomorphism between the (equivalence classes of
the) space of graphs with a prescribed length to each edge and the
decorated moduli space. It follows from the definitions above that
the set of critical curves of any Strebel differential is a graph.
Given the critical graph, we assign a length to each edge of the
critical graph by the integral along the edge
$\int_{z_1}^{z_2}\sqrt{q}$, where the integral is performed in the
orientation for which it is positive (the integrand is real since
the curve is horizontal).  Note that $z_1$ and $z_2$ are zeros of
$q$, and the integral is carried over the critical curve, although
the result for any homotopic curve is the same (correctly dealing
with the branch cuts).  Using Strebel's theorem, this gives a
mapping from the decorated moduli space to graphs with edge
lengths. The opposite mapping, gluing a graph with given edge
lengths to form a specific Riemann surface, is a bit more
intricate and is described in the references above.

\subsection{Cell decompositions}

In this subsection we show how the above results give rise to cell
decompositions of $X$ and of the decorated moduli space.

\begin{figure}[!ht]
\begin{center}
\input{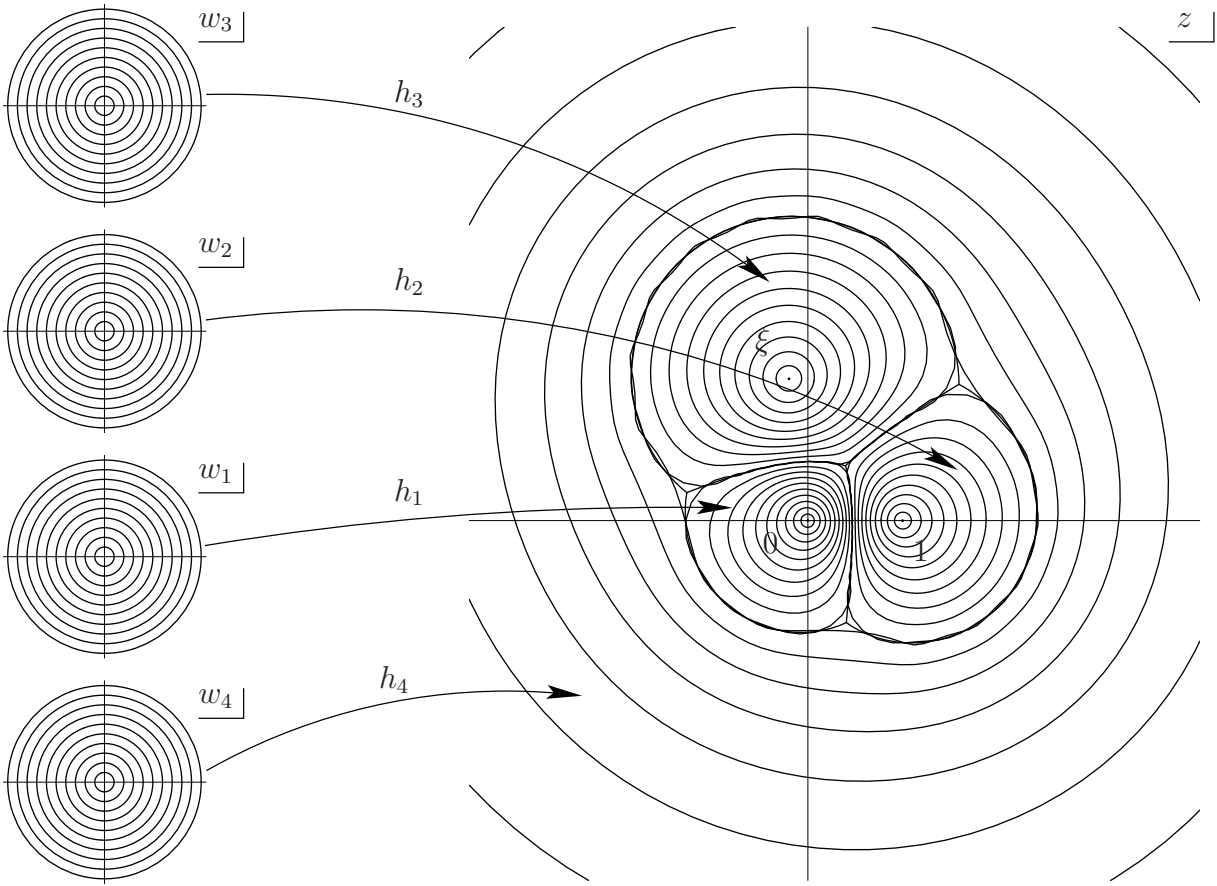}
\caption{\footnotesize{The horizontal curves for a four-punctured
sphere. Each domain is conformally a disk. The
critical graph is easily seen. This figure was produced by a
numerical analysis which is described in \cite{Moeller:2004yy}. }}
\label{conformal_f}
\end{center}
\end{figure}

Firstly, the unique Strebel differential on a specific marked
Riemann surface $X$ induces a cell decomposition of $X$ in the
following manner. One collects all vertices of the critical graph
(which are the zeros of $q$) in the 0-cell part of the complex.
The 1-cells are the edges of the critical graph, and finally, the
2-cells are the domains around the poles, which are foliated by
the closed simple horizontal curves. This gives a cell
decomposition because the domains around the poles are conformal
to disks. In particular, one may define a new holomorphic
coordinate $w_i$ around each $x_i$ such that the differential in
the whole domain foliated by the closed simple horizontal curves
is given exactly by
\begin{equation}
q=-\frac{p_i^2}{(2\pi)^2 w_i^2}dw_i^2.
\end{equation}
Of course, the Euler formula
$v-e+f=2-2g$ is satisfied,
where $v$ is the number of 0-cells, $e$ is the number of 1-cells and $f$
is the number of 2-cells. Figure \ref{conformal_f},
copied from \cite{Moeller:2004yy}, shows this decomposition for
the case of four punctures on the sphere (conformally mapped to the
plane with the punctures at $0,1,\xi,\infty$).

Secondly, the isomorphism of the decorated moduli space with the
metric graphs induces a cell decomposition of the decorated moduli
space. It is clear that any vertex of order higher than $3$ in the
critical graph can be split in various channels until the graph
has only vertices of order $3$. This can be done without
intersecting other edges or affecting the genus. Hence, a generic
Riemann surface is mapped to a Strebel differential which has at
most simple zeros. Consequently, these trivalent graphs sweep out
a top dimensional cell of the decorated moduli space. Riemann
surfaces which map to graphs with higher order vertices (coming
from at least two zeros that have merged) are part of some lower
dimensional cell in the complex. It can be shown that this
description is compatible with the continuity notion on the
decorated moduli space, and hence can be viewed as a cell
decomposition. This result has had many applications both in
mathematics and in physics. Some are summarized in the references
of \cite{Gopakumars} (see especially the Kontsevich theory
\cite{Konts}).

\subsection{Schwinger parametrization of Feynman diagrams}

\begin{figure}[tbp]
\begin{center}
\includegraphics{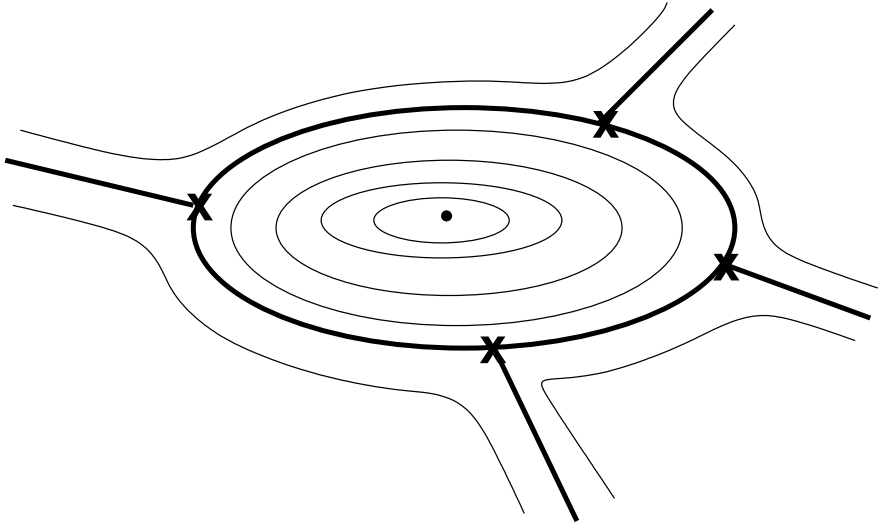} \caption{\footnotesize{(borrowed from
\cite{Gopakumars}) A characteristic ring
domain in the vicinity of a double pole (marked with a dot). The
non-closed horizontal trajectories are shown by thick lines. These
begin and end at zeros marked by a cross. }} \label{ring domain}
\end{center}
\end{figure}

As described in the introduction, we wish to map the Feynman
diagrams of a free gauge theory to closed string diagrams, namely
to punctured Riemann surfaces.  It is quite clear that the
critical graph itself is not similar in any way to the typical
Feynman graph of a free gauge theory (where we put composite
operators such as $\tr(F_{\mu \nu}^n)$ as insertions).  Indeed, in
the vicinity of some second order pole the critical graph would
generically look like Figure \ref{ring domain}. This is not
adequate to describe a subgraph of a free gauge theory diagram;
rather, this seems to describe vacuum-vacuum transitions in some
theory with a cubic interaction term and elementary fields in the
adjoint representation (such as the Kontsevich matrix model
\cite{Konts}). To figure out how to recast free gauge theory
graphs in this form we make a detour in this subsection to
introduce the Schwinger parametrization of Feynman diagrams.

For simplicity, we will discuss only correlation functions of
operators involving adjoint scalar fields $\Phi$, of the form
$\vev{\tr(\Phi^{i_1}(x_1)) \tr(\Phi^{i_2}(x_2)) \cdots
\tr(\Phi^{i_n}(x_n))}$.  A Euclidean propagator with momentum $p$
of a scalar field of mass $m$ can be rewritten as an integral over
the Schwinger parameter $\tau$ of the propagator,
\begin{equation}\label{schwinger}
\frac{1}{p^2+m^2}=\int^{\infty}_{0}d\tau e^{-\tau(p^2+m^2)}.
\end{equation}
It is usually easier to compute free field theory correlation functions
in position space
rather than in momentum space, since there are no integrations required
in position space. The propagator in position space from $x_i$ to
$x_j$ is the Fourier transform of (\ref{schwinger}).
For the special case of massless scalar fields in four dimensions
(which we will focus on for the rest of this paper, though the
generalization to any other fields should be straightforward),
it is simply given by (up to a constant)
\begin{equation}\label{schwingerp}
\frac{4}{(x_i-x_j)^2}=\int^{\infty}_{0}d\sigma
e^{-\sigma(x_i-x_j)^2/4},
\end{equation}
where $\sigma \equiv 1/\tau$ is the inverse Schwinger parameter.
In order to avoid divergences,
we do not consider lines that begin and end on the same vertex (so
our operators will be normal ordered). A general Feynman diagram
is given by a product of such factors over all the propagators in
the diagram -- if the $k$'th propagator in the Feynman diagram
($k=1,\cdots,E$) connects
$x_{i_k}$ with $x_{j_k}$, the amplitude is given (up to a constant) by
\begin{equation}\label{Pos_1}
\int \prod^E_{k=1} d\s_k e^{-(x_{i_k}-x_{j_k})^2\s_k/4}.
\end{equation}

\begin{figure}[tbp]
\begin{center}
\includegraphics[height=7cm]{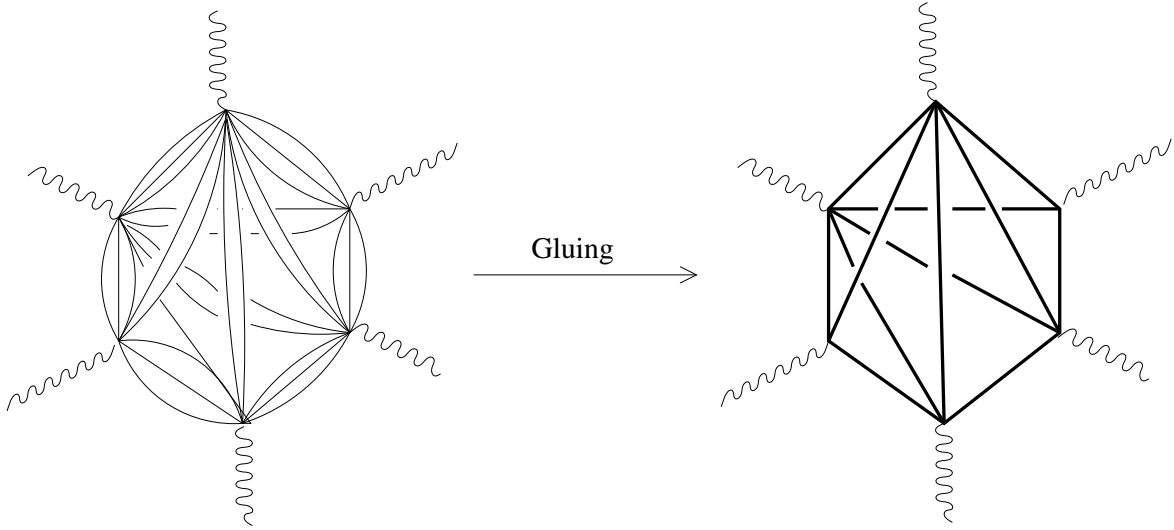}
\caption{\footnotesize{(borrowed from \cite{Gopakumars})
A specific (planar) 6-point diagram
is reduced to a skeleton graph by collapsing color loops which are
homotopic to a line. }}
\label{skeleton}
\end{center}
\end{figure}

As noted in \cite{Gopakumars},
any two homotopic edges in the Feynman diagram can be combined
into an integral over an effective Schwinger parameter satisfying
\begin{equation}\label{join}
\frac{1}{\tau_{eff}}=\frac1{\tau_1}+\frac1{\tau_2} \hspace{1em}
\text{or}\hspace{1em} \sigma_{eff}=\sigma_1+\sigma_2.
\end{equation}
This is the well-known analogy to electrical networks (it is most
easily understood by inspection of (\ref{Pos_1}) where the
integrand manifestly depends only on the sum of $\sigma$'s for
homotopic lines).
Thus, we can collapse all homotopic lines to a single line,
but with a modified dependence on the effective Schwinger parameters
(following from (\ref{join})).
Such reduced diagrams will be named skeleton
graphs. Note that the notion of homotopy is well defined
after the diagram is drawn on some Riemann surface, or
alternatively is given in double line notation where loops that
respect the color flow are apparent. Strictly speaking, in a
gauge theory with adjoint matter we can also combine lines in
this way for non-homotopic lines between the same pair of points,
but this is not useful for reasons that will become clear below.
The resulting formula for the amplitude, if we have ${\tilde E}$
reduced edges, with the $r$'th edge (connecting $x_{i_r}$ and
$x_{j_r}$) coming from joining together
$m_r$ homotopic lines, such that its Schwinger parameter is
$\tilde{\sigma}_r=\sum_{\mu_r=1}^{m_r} \sigma_{\mu_r}$, is
(up to a constant depending on the $m_r$)
\begin{equation}\label{Pos_2}
\int \prod^{\tilde E}_{r=1} d{\tilde \s}_r {\tilde \s}_r^{m_r-1}
e^{-(x_{i_r}-x_{j_r})^2 {\tilde \s}_r/4}.
\end{equation}
Pictorially, this procedure looks like (it is not
easily drawn for some non-planar diagrams) Figure \ref{skeleton}.
%

\subsection{The mapping of a free field theory graph to a
closed string amplitude}

Now we are ready to describe the mapping from a free field theory
graph to a closed string graph, by recasting the integrals over
the Schwinger parameters into integrals over the decorated moduli
space of Riemann surfaces.
The ``skeleton reduction'' procedure reviewed above
suggests that the generic field
theory Feynman diagram may be viewed as a triangulation of the appropriate
surface. Since the critical graphs were generically trivalent,
it is natural to conjecture that the rearrangement
into integrals over the moduli space of Riemann surfaces should go
through graph duality. In particular, the field theory insertions which
are the vertices of the Feynman graph should map
to poles of second order in the differential, which are faces of the
critical graph, and the faces of the Feynman graph should map
to the zeros of the differential. Each edge of the Feynman diagram,
associated with an effective Schwinger parameter $\tau_i$, is mapped by the
graph duality to an edge of the critical graph of length $l_i$.
Gopakumar's suggestion for the mapping was to identify $l_i$
with $1/\tau_i = \sigma_i$. This maps the integration over
Schwinger parameters in the amplitude (\ref{Pos_2}) to an integration
over the space of graphs with edge lengths, which can then be
mapped to an integration over the decorated moduli space. This
can then be identified as a closed string amplitude (after the integration
over the parameters $p_i$ associated with the circumferences of the holes), and
the procedure described above can be used to write down the integrand
(the worldsheet correlation function) in this amplitude.
%
%

It is obvious from the construction that the parameters on both
sides match. As we noted, by skeleton reduction the most generic
field theory diagram reduces to some triangulation of the Riemann
surface it lives on, so the dual graph will consist of trivalent
vertices. By Euler's theorem (applied to the dual graph),
$v-e+f=2-2g$ where $v$, $e$, $f$ are the numbers of vertices,
edges, and faces respectively. For the case of trivalent vertices
$e=\frac{3}{2}v$. In the dual graph, $f$ is the number of field
theory insertions. So, we denote $f\equiv n$ and we obtain that
the number of lengths that we integrate over is
\begin{equation}
e=6g-6+3n=6g-6+2n+n.
\end{equation}
As is well known, $6g-6+2n$ is the number of real moduli for
a Riemann surface of genus $g$ with $n$ insertions (note that for
$g\leqslant 1$ one has to saturate with the right number of vertex
operators needed by the order of the conformal Killing group to use the above
formula). The additional $n$ that we obtained corresponds to the moduli $p_i$
of the decoration, $\mathbb{R}_+^n$.

It is clear that the worldsheet theory which is dual to a free
gauge theory must have some strange properties. For instance, a
given correlation function in a free gauge theory only gets
contribution from a finite number of Feynman graphs whose genus is
bounded by $g \leq g_0$, so the corresponding worldsheet
correlation functions must vanish on all Riemann surfaces of genus
$g>g_0$. This is a generic property of any mapping of free field
Feynman diagrams to Riemann surfaces. However, the specific
mapping suggested by Gopakumar also has some other strange
properties. As we noted, generic Feynman diagrams (with many
edges) will map to generic points on the decorated moduli space.
However, in specific correlation functions one could get
contributions also from less generic diagrams that have less
edges. On the dual graph side such diagrams correspond to joining
together several zeros of the Strebel differential, so they are
localized on subspaces of the decorated moduli space.
In particular, for some correlation functions the only contributions
come from non-generic graphs. For example, consider the correlation
function (involving a single adjoint scalar field)
$\vev{\tr(\Phi^2(x_1)) \tr(\Phi^2(x_2)) \cdots \tr(\Phi^2(x_n))}$.
The only connected Feynman diagrams here are circular graphs which
have the topology of a
sphere, $n$ vertices and $n$ edges, so the dual graph has $f=e=n$ and
$v=2$. This Feynman diagram is mapped by Gopakumar's mapping to an
$n$-dimensional subspace of the decorated moduli space, which is
(for $n \geq 3$) of
dimension $-6+3n$. Moreover, for $n>6$ this subspace has a lower
dimension than the moduli space of marked spheres itself (of
dimension $2n-6$), so
even after we integrate over the hole sizes $p_i$, the corresponding
amplitude will vanish for generic insertion positions, and will only
be non-vanishing for specific insertion positions. This is quite
surprising, since one would think that the correlation functions on
the worldsheet should be smooth functions of the insertion positions.
We will encounter several more examples of this phenomenon below.
Even generic correlation functions, for which there are contributions
from generic Feynman graphs, could have contributions also from graphs
with less edges, and these contributions would give non-smooth
worldsheet correlation functions as described above.
%

It is clear that an identification of the string theory dual of
free gauge theories, of the type described above, can
straightforwardly be generalized to any order in perturbation
theory, simply by identifying the string theory duals of the
interaction vertices in the space-time theory and adding them to
the worldsheet action. Obviously, at any finite coupling a
specific correlation function gets contributions involving any
number of interaction vertices, so the strange properties
described above would disappear. This may suggest that the limit
of taking the gauge coupling constant to zero is a singular limit,
but one can still hope that it might be possible to understand the
corresponding worldsheet theory.

Note that in our definition of free field theory correlation
functions above we assumed that the operators are all
normal-ordered; this is necessary for all correlation functions to
be finite (and for preserving conformal invariance in the case of
four dimensional gauge theories). However, the mapping described
above from decorated Riemann surfaces to Feynman graphs gives
generic Feynman graphs, which can include also self-contractions
(propagators from a vertex to itself), as long as these
self-contractions are toplogically non-trivial (they go around
another insertion or around a non-contractible curve of the
Riemann surface). Again, this will result in large regions of the
decorated moduli space on which the correlation functions on the
worldsheet will vanish (since the corresponding field theory
Feynman diagrams are set to zero). We will see explicit examples
of this below.

Finally, we wish to describe another peculiar feature of the
mapping described above from Feynman diagrams to worldsheets.
It is natural to expect the worldsheet theory to respect all the
symmetries of the space-time theory. In particular, the worldsheet
vertex operators should transform under the Poincar\'e group in
the same way as the corresponding space-time operators.
And indeed, the mapping described above satisfies this property,
since the Schwinger parameters (which can be thought of as proper
times along the propagators) are Poincar\'e-invariant, so a
Poincar\'e transformation does not change the (decorated) moduli.
Next, let us assume that the space-time theory is invariant under
scaling transformations $x \to \alpha x$ (as is the case for a
free gauge theory with massless matter fields).
Under a scaling transformation the Schwinger parameter $\tau$
transforms as $\tau\to \alpha^2 \tau$. This means that the hole
radii $p_i$ are not invariant under dilatations, because the hole
radii are given by the appropriate circumferences (which are
linear combinations of $\sigma_j$'s). However, note that any
Strebel differential, multiplied by a positive constant, is still
a Strebel differential on the same Riemann surface. Thus, the
scaling transformation in space-time will act on the decorated
moduli space just by multiplying the Strebel differential by a
constant, and in particular the Riemann surface moduli are
invariant (note that this implies that these moduli depend only on
ratios of Schwinger parameters). Hence, the worldsheet vertex
operators are expected to have well defined scaling
transformations. Now, let us consider special conformal
transformations (which are symmetries of four dimensional free
gauge theories with massless matter fields). The special conformal
transformation generated by the inversion $x^{\mu} \to
x^{\mu}/x^2$ acts in a more involved way on the Schwinger
parameters. The propagator (for massless four dimensional scalar
fields) from $x_i$ to $x_j$ is given by the expression $\int
d\sigma e^{-\sigma (x_i-x_j)^2/4}$, so in order for it to
transform properly under inversion we need to take
$\sigma \rightarrow x_i^2x_j^2\sigma$.
%
The formulas relating the Schwinger length
parameters to the Riemann surface moduli are not
invariant under such a transformation (since each Schwinger length
transforms in a different way). Thus, the special conformal transformations
act non-trivially on the Riemann surface moduli and not just on the
vertex operators, and we do not expect to have
well defined transformation laws of worldsheet vertex operators
under special conformal transformations. This will be shown explicitly
below. Note that this implies that the full conformal group is not
realized as a global symmetry of the worldsheet theory, but only its
subgroup including the Poincar\'e and scaling transformations.

\section{On Worldsheet and Space-time OPEs}\label{OPEsect}

\def\vev#1{\langle #1 \rangle}

In this section we discuss our expectations about the worldsheet
OPEs in the string theory which is dual to a free large $N$ gauge
theory, and their possible relations to the space-time OPEs. For
simplicity we write down formulas which are correct for $d=4$, but
the generalization to other dimensions is straightforward.

The space-time OPEs in free gauge theories are rather trivial. If
we just had a theory of (say) a free scalar field $\phi$, the OPE
of $\phi(x)$ with $\phi(y)$ would contain a single singular term
of the form $I/|x-y|^2$ (where $I$ is the identity operator), and
non-singular terms of the form $(x-y)^n (\phi \partial^n \phi)(y)$
for $n=0,1,\cdots$. In a gauge theory the OPE is slightly less
trivial because we need to look only at gauge-invariant operators.
For example, consider a scalar field $\Phi$ in the adjoint
representation of $SU(N)$. The gauge-invariant operators we can
make out of this field include $\tr(\Phi^n(x))$ for $n \geq 2$ (in
the large $N$ limit all these operators are independent), traces
of products of $\Phi$'s which include derivatives, and multi-trace operators
which are products of these traces.  The OPE of an operator
$\tr(\Phi^{n_1}(x))$ with $\tr(\Phi^{n_2}(y))$ takes the form (as
$x\to y$)
\begin{equation}\label{stope}
\tr(\Phi^{n_1}(x)) \tr(\Phi^{n_2}(y)) \sim
\sum_{n=|n_1-n_2|}^{n_1+n_2} C_{n_1,n_2,n} |x-y|^{n-n_1-n_2}
\tr(\Phi^n(y)) + {\rm other\ operators},
\end{equation}
where the ``other operators'' include both operators with
derivatives and multi-trace operators. In the large $N$ limit, the
coefficients $C_{n_1,n_2,n}$ have an expansion in powers of
$1/N^2$, corresponding to the genus of the free-field diagrams
contributing to the three-point functions represented in
(\ref{stope}) (when they are written in 't Hooft's double-line
notation \cite{'tHooft:1973jz}). Note that the space-time
dependence of the OPE coefficients is determined by the dimensions
of the participating operators, since the free gauge theory is
scale-invariant. 

As discussed above, we expect that the free large $N$ gauge theory
should have a string theory dual with a string coupling scaling as
$1/N$, in which each of the single-trace operators of the gauge
theory ${\cal O}(x)$ should be represented by a vertex operator on
the worldsheet $V_{\cal O}(x;z)$ (we will use $x,y$ to denote
space-time positions and $z,w$ to denote worldsheet positions). We
do not know much about what this worldsheet theory is for specific
free gauge theories, but we expect that we should be able to put
it into a conformal gauge, in which the space-time correlation
functions are equal to correlation functions of integrated vertex
operators $\int d^2z V_{\cal O}(x;z)$. In this gauge, all physical
worldsheet vertex operators should have worldsheet scaling
dimension $(h,{\bar h})=(1,1)$. In particular, we expect each
single-trace gauge-invariant operator ${\cal O}(x)=\tr(\cdots(x))$
of the space-time theory (with or without derivatives) to
correspond to a vertex operator of this dimension.

A natural object to consider in the worldsheet theory is the
worldsheet OPE; we expect that as $z\to w$, the product $V_{{\cal
O}_i}(x;z) V_{{\cal O}_j}(y;w)$ of two physical vertex operators
should have an expansion in local worldsheet operators of the form
\begin{equation}\label{wsope}
V_{{\cal O}_i}(x;z) V_{{\cal O}_j}(y;w) \sim \sum_k
(z-w)^{h_k-2} ({\bar z}-{\bar w})^{{\bar h}_k-2} C_{ijk}(x,y)
V_k(z),
\end{equation}
where the operators $V_k$ which appear have worldsheet scaling
dimensions $(h_k,{\bar h}_k)$. Obviously, there is no reason why
the operators on the right-hand side of (\ref{wsope}) should all
be physical operators; we expect the OPE in general to include
both physical operators (of dimension $(1,1)$, leading to an OPE
coefficient proportional to $1/|z-w|^2$) and other operators
with various dimensions.

As in any theory, the OPE coefficients $C_{ijk}$ are related to
the three-point functions of the operators $V_{{\cal O}_i}
V_{{\cal O}_j} V_k$ (the two are proportional to each other in a
basis where the two-point functions of the $V_k$'s are diagonal).
In particular, they should be non-vanishing whenever the
three-point function is non-vanishing. Thus, if we have a
non-vanishing three-point function of three operators in
space-time (as implied, for instance, by (\ref{stope})), we know
that the OPE coefficient of the corresponding physical worldsheet
operators should also be non-vanishing. In this case, where $V_k$
corresponds to a physical operator, the operators $V_k$ are
labeled by a space-time position $u$, such that the sum in
(\ref{wsope}) is really an integral, and the OPE coefficients
$C_{ijk}(x,y)$ are a product of powers\footnote{In general,
contact terms could also appear.} of $|x-y|$, $|x-u|$ and $|y-u|$
(we will discuss the precise form in more detail below). It is
natural to expect that also the non-physical vertex operators
appearing in the OPE should have a space-time interpretation, but
it is not completely clear that only local space-time operators
should appear. Note that in this discussion (as above) we assumed
that the free field theory lives on Minkowski space; obviously for
a theory living (for instance) on a compact space-time, we could
expand all the gauge-invariant operators in Kaluza-Klein modes on
the compact space-time and the OPE discussed above would be
discrete rather than continuous.

In the previous paragraph we discussed one source of a continuum
appearing in the OPE, related to the non-compactness of the
Minkowski space-time which the field theory lives on.  In some
sense this continuum is fake, because if we label the operators by
their space-time momentum $p$ rather than their position $x$,
defining ${\tilde V}_{{\cal O}_i}(p;w) = \int d^4x e^{ipx}
V_{{\cal O}_i}(x;w)$, then the momentum of the operator appearing
on the right-hand side of a product of ${\tilde V}$'s would be
determined by momentum conservation and this continuum would
disappear. However, in string theories which provide holographic
descriptions of field theories there is another source of a
continuum which does not disappear even in momentum space. When we
have a ``bulk space-time'' interpretation of these string
theories, for instance (for conformal field theories) as sigma
models on anti-de Sitter space, this continuum arises from the
non-compactness of the ``radial direction'' which is
holographically related to the energy scale of the dual field
theory. For physical vertex operators, corresponding to on-shell
fields in this ``bulk space-time'' (but not in the original field
theory), the dependence on this ``radial direction'' is determined
by the momentum in the field theory directions; but we expect to
have also unphysical vertex operators, loosely corresponding to
off-shell fields in the ``bulk space-time'', which can have an
arbitrary dependence on the ``radial direction'', leading to a
continuum of operators with various worldsheet dimensions.

In general we would expect all the operators with the appropriate
quantum numbers to appear in the worldsheet OPE, so we would
expect the right-hand side of (\ref{wsope}) to include a continuum
of dimensions (note that this is different from the continuum
coming from the non-compact space-time directions for the physical
vertex operators discussed above, for which all the operators had
the same worldsheet dimension). The existence of such a continuum
is expected on general grounds, and it explicitly appears in the
worldsheet OPE whenever the worldsheet theory holographically dual
to a field theory is known, for example in sigma models on $AdS_3$
\cite{Teschner,MaldacenaOoguri} or 
$SL(2)/U(1)$ \cite{GiveonKutasov}. One technical reason which
necessitates the appearance of such a continuum is that the
space-time theory has non-vanishing planar two-point functions
(for instance $\vev{\tr(\Phi^n(x)) \tr(\Phi^n(y))}$), while in
standard string theories all two-point functions on the sphere
vanish; this is consistent because the operators of the worldsheet
theory are generally labeled by a continuous parameter $j$
(determining their dependence on the ``radial position'') such
that their two-point functions behave as $V_{j_1} V_{j_2} \propto
\delta(j_1-j_2)$, and the infinity in the delta function cancels
the infinity from the volume of the symmetry group of the sphere
with two punctures \cite{Kutasov:1999xu}.  Another reason why a continuum must
always appear is that, as discussed above, physical vertex
operators on the right-hand side of (\ref{wsope}) lead to an OPE
behaving as $1/|z-w|^2$. If we look at an $n$-point function
involving the operators $V_{{\cal O}_i}(z)$ and $V_{{\cal O}_j}(w)$ with
$n > 3$, the position $z$ should be integrated over, and such a behavior
would lead to a divergent integral in the region $z \sim w$
(unless the correlation function of $V_k$ with the
other operators in the correlator vanishes, which we do not expect
to happen generically). On the other hand, the position space
space-time correlation functions in the free field theory
(assuming the operators are normal-ordered) are all finite, and we
do not expect any divergences to appear. The resolution is that in
general the physical vertex operator on the right-hand side would
appear as part of a continuum of operators $V_{\alpha}$ of
worldsheet scaling dimension $2+\alpha$, and the correlator would
behave as $\int d^2z \int d\alpha C_{\alpha} / |z-w|^{2-\alpha}$
rather than as $\int d^2z /|z-w|^2$; we expect that the form of
$C_{\alpha}$ should be such that no divergence appears as $z\to
w$, and again this is confirmed in the cases of $AdS_3$ and
$SL(2)/U(1)$ where one can explicitly analyze the worldsheet
theory.

So far we have separately discussed our expectations concerning
the space-time and worldsheet OPEs; we now ask if there is any
relation between them. As we mentioned above, in general there is
no such relation. We expect that any operator appearing in the
space-time OPE should also appear as a physical operator in the
worldsheet OPE, but in general many more operators (physical or
unphysical) should also appear in the worldsheet OPE, which have
no clear interpretation in the space-time OPE. However, it is
still natural to expect that the worldsheet OPE should have some
space-time interpretation, since when two points on the worldsheet
come together, one expects their images in space-time to come
together as well. Thus, we expect that the leading terms in the 
worldsheet OPE will
depend strongly on the space-time positions $x,y$ labeling the
worldsheet operators. In some cases we expect contact terms
(behaving as a derivative of a delta function of $x-y$) to appear,
while in other cases, when the OPE includes a worldsheet operator
which can be interpreted as a local space-time operator at a
position $u$, we expect the coefficients to diverge as $x$, $y$
and $u$ come together.  Again, these expectations are confirmed in
the known case of $AdS_3$, and we will see that they are also
confirmed by the OPEs we will find below.

As discussed above, we expect many of the operators appearing on
the right-hand side of the OPE (\ref{wsope}) to behave as local
operators in space-time (even though, when they are not physical,
they do not map to actual space-time operators). Then, a
specific term in the OPE takes the form
\begin{equation}\label{wsopest}
V_{{\cal O}_i}(x;z) V_{{\cal O}_j}(y;w) \sim \int d^du
(z-w)^{h_k-2} ({\bar z}-{\bar w})^{{\bar h}_k-2} C_{ijk}(x,y,u)
V_k(u;z).
\end{equation}
It is natural to expect that the operators $V_k$ should scale
under space-time scaling transformations with a fixed scaling
dimension $\Delta_k$ (for example, the physical vertex operator on
the worldsheet corresponding to $\tr(\Phi^n(x))$ should scale
under space-time scaling transformations as an operator of
dimension $n$). Naively we can then use the conformal symmetry to
determine the space-time dependence of $C_{ijk}$; for a scalar
(in space-time) operator we would obtain
\begin{equation}\label{confconstraint}
C_{ijk}(x,y,u) = C_{ijk} |x-y|^{d-\Delta_k-\Delta_i-\Delta_j}
|x-u|^{\Delta_j-\Delta_i-d+\Delta_k}
|y-u|^{\Delta_i-\Delta_j-d+\Delta_k}.
\end{equation}
However, as mentioned in the previous section, the
translation of \cite{Gopakumars} from the space-time to the
worldsheet does not preserve the full conformal group, but just
the Poincar\'e and scaling symmetries. These do not constrain the
coefficients to the form (\ref{confconstraint}), but more general
forms are also allowed (of overall space-time scaling dimension $\Delta_i+
\Delta_j+d-\Delta_k$). In our analysis of four-point functions
below, we will see that indeed more general forms do arise which
are not consistent with the full conformal symmetry group.

\section{Sphere Diagrams}
\label{Sphere}
We begin our analysis of the worldsheet properties of the string
theory dual to free gauge theories with planar (sphere) amplitudes. The
analysis of general amplitudes is complicated already
at this level, even for correlators with a small number of
insertions, because
finding the Strebel differential
involves solving implicit integral equations as well as
simple algebraic ones. We will show that one can learn interesting
lessons by considering very simple special diagrams.
We begin by discussing the simplest example, of
three-point functions on the sphere, to illustrate how the general
formalism works. Then, we calculate the simplest
four and five-point diagrams one can construct.
The analysis of more general cases (on which we were not able to
make as much progress) may be found in appendix \ref{SphereApp}.
\subsection{Three-point function on the sphere}
The simplest non-trivial correlation function in a standard string theory
is the three-point function on the sphere (for string duals to field theories
there are also planar two-point functions, but we will not discuss them
here). In this case there is no moduli space, since we can use the worldsheet
conformal symmetry
to fix the three insertions on the worldsheet to $z=0,1,\infty$. Thus, the
worldsheet three-point function is exactly the same as the space-time
three-point function. However, as described above, in Gopakumar's formalism
we work on the decorated moduli space, and translate the Feynman diagrams
into functions on the decorated moduli space (whose integral over the
circumferences $p_i$ gives the closed string correlator). For a three-point
function this decorated moduli space is three dimensional.

For simplicity, we consider only simple scalar correlation
functions (in four dimensional gauge theories) of the form
\be
\langle \tr(\Phi^{J_1}(x_1)) \tr(\Phi^{J_2}(x_2))
\tr(\Phi^{J_3}(x_3))\rangle_{S^2}.
\ee
The most general Strebel differential for this problem is\footnote{The
essential part of the analysis appears in \cite{Mulase:98}.}:
\begin{equation}
\label{eq: strebel differential on P1} q = -\frac{1}{4\pi^2}\left(
a \left(\frac{dz}{z}\right)^2 + b \left(\frac{dz}{1-z}\right)^2 +
c \left(\frac{dz}{z(1-z)}\right)^2 \right),
\end{equation}
where the parameters $a,b,c$ are related to the residues $p_i$ by
\begin{equation}
a = \frac{1}{2} \left( p_0 ^2 + p_{\infty} ^2 - p_1 ^2\right),\qquad
b = \frac{1}{2} \left( p_1 ^2 + p_{\infty} ^2 - p_0 ^2\right),\qquad
c = \frac{1}{2} \left( p_0 ^2 + p_1 ^2 - p_{\infty} ^2\right).
\end{equation}
%
This equation gives us a simple explicit formula for the Strebel
differential as a function of the coordinates of the decorated
moduli space.
There are three
different types of possible critical graphs, depending on where we
are in the decorated moduli space. The three cases are
characterized by the sign of $\Delta \equiv ab+ac+bc$ :
\begin{figure}[htbp]
\begin{center}
$\begin{array}{c@{\hspace{.3in}}c@{\hspace{.3in}}c}
\epsfig{file=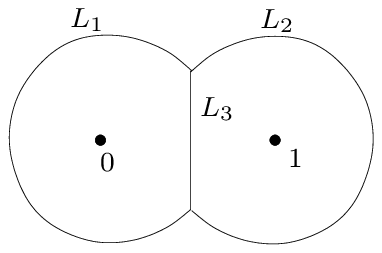, width=1.6in} & \epsfig{file=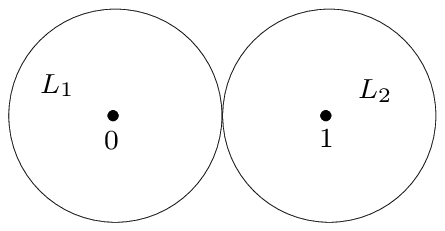,
width=1.9in} & \epsfig{file=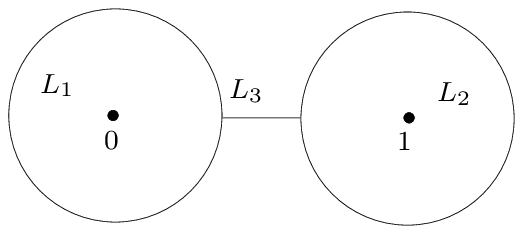, width=2.1in}
 \\ [0.2cm]
\mbox{\bf (a)} & \mbox{\bf (b)} & \mbox{\bf (c)}
\end{array}$
\caption{The three possibilities for the topology of the critical curves of the
Strebel differential : {\bf (a)} $\Delta>0$,  {\bf (b)}
$\Delta=0$, {\bf (c)} $\Delta<0$.} \label{regs3Point}
\end{center}
\end{figure}
%
%
\begin{itemize}
\item $\Delta>0$
: A representative critical graph for this case is drawn in Figure
\ref{regs3Point}(a). The relation between the edge lengths $L_i$
of the critical graph and the circumferences $p_i$ is given by \be
\label{dic1} L_1=\half(p_0+p_\infty-p_1),\qquad
L_2=\half(-p_0+p_\infty+p_1),\qquad L_3=\half(p_0-p_\infty+p_1).
\ee The quantity $\Delta$ for this graph
is equal to $16L_1L_2L_3(L_1+L_2+L_3)$, which is indeed positive.
\item $\Delta=0$
: This is a degenerate case (see Figure \ref{regs3Point}(b) for an
example) in which the two zeros of the Strebel differential have
joined together, and one of the edges has degenerated; in the case
depicted in the figure (the other cases are related by
permutations of the points) the edge lengths are given by: \be
L_1=p_0,\quad L_2=p_1,\quad  p_{\infty}=p_0+p_1.\ee
\item $\Delta<0$
: In this case the topology of the critical graph is given by
Figure \ref{regs3Point}(c) (up to permutations of the three
points), and the edge lengths are given by: \be L_1=p_0,\qquad
L_2=p_1, \qquad L_3=\half(p_\infty-p_1-p_0). \ee
\end{itemize}
In Figure \ref{Regions} we show where each of the different possible
three-point critical graphs appears in the decorated moduli space.
\begin{figure}[htbp]
\begin{center}
\input{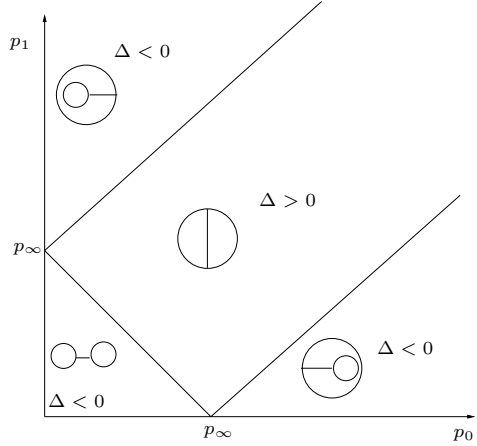}
\caption{The critical graphs in the
different regions in the decorated moduli space. We exhibit
a slice with constant $p_\infty$. On the lines
$p_0+p_1=p_\infty$ and $p_0=p_1\pm p_\infty$ we have the
interpolating degenerate diagram, such as Figure \ref{regs3Point}(b).}
\label{Regions}
\end{center}
\end{figure}

After writing down the Strebel differentials we can translate any
given three-point function in the gauge theory to the string
theory language, writing it as an integral over the decorated
moduli space. In each region we have a different critical graph
corresponding to a different dual field theory graph, so a given
Feynman diagram will contribute only in one of the three regions
described above. We now discuss the translation in each of the
three regions :
\begin{itemize}
\item $\Delta>0$
\begin{figure}[htbp]
\begin{center}
$\begin{array}{c@{\hspace{.5in}}c@{\hspace{.5in}}c}
\epsfig{file=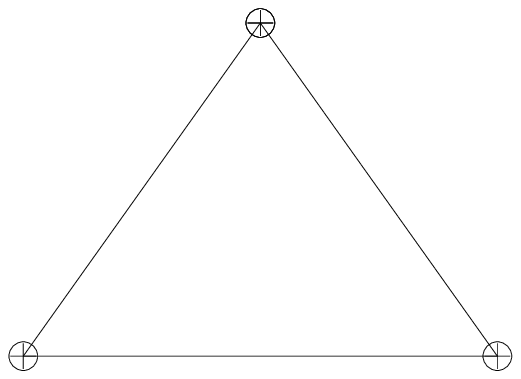, width=1.4in} & \epsfig{file=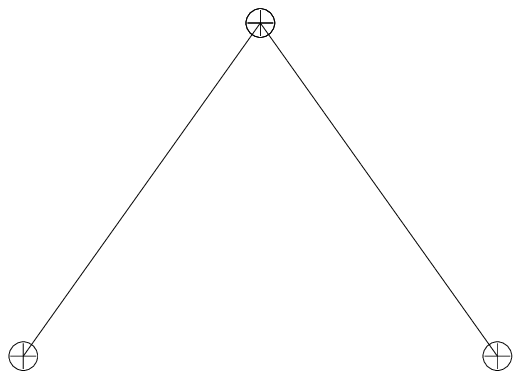,
width=1.4in} & \epsfig{file=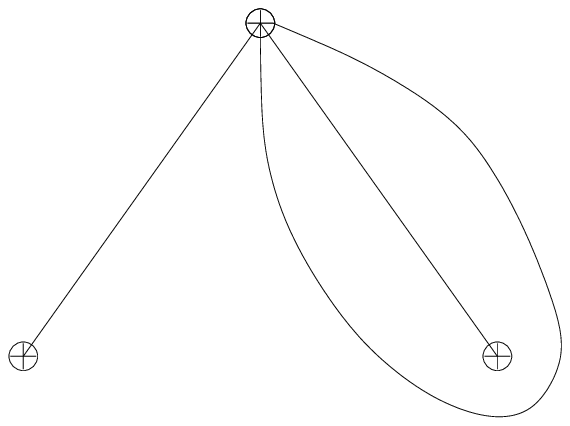, width=1.4in}
 \\ [0.2cm]
\mbox{\bf (a)} & \mbox{\bf (b)} & \mbox{\bf (c)}
\end{array}$
\caption{The possible gauge theory graphs : {\bf (a)} $\Delta>0$,
{\bf (b)} $\Delta=0$, {\bf (c)} $\Delta<0$. In this paper we
denote composite operators by a circle with a cross in it. }
\label{regs3PointF}
\end{center}
\end{figure}
: The gauge theory diagram for this case is given in Figure
\ref{regs3PointF}(a) (which is the dual graph of Figure
\ref{regs3Point}(a)). As a specific example, we discuss the
correlation function $\langle \tr(\Phi^2(x_1)) \tr(\Phi^2(x_2))
\tr(\Phi^2(x_3))\rangle_{S^2}$ for which this is the only
contributing diagram. When written in terms of the inverse
Schwinger parameters $\sigma_i$, which we identify with the
lengths $L_i$ of the critical graph, the diagram is given by
%
\begin{align}
\label{posDelta}
G&=A\int d\sigma_1 d\sigma_2 d\sigma_3
 e^{-\frac{(x_2-x_1)^2\sigma_1}{4}}e^{-\frac{(x_2-x_3)^2 \sigma_2}{4}}
e^{-\frac{(x_3-x_1)^2\sigma_3}{4}}=\cr
&= \frac{64A}{(x_2-x_1)^2(x_3-x_1)^2(x_2-x_3)^2},
\end{align}
where $A$ is a numerical factor.
To write this expression as an integral over the circumferences we use
the dictionary (\ref{dic1}). The integral then becomes
\be
G=\half A \int_0^\infty dp_0\int_0^\infty
dp_1\int_{|p_0-p_1|}^{p_0+p_1} dp_\infty
e^{-\frac{p_0}{8}((x_2-x_1)^2-(x_2-x_3)^2+(x_3-x_1)^2)}\times\nonumber \\
e^{-\frac{p_1}{8}(-(x_2-x_1)^2+(x_2-x_3)^2+(x_3-x_1)^2)}
e^{-\frac{p_\infty}{8}((x_2-x_1)^2+(x_2-x_3)^2-(x_3-x_1)^2)},
\ee
which is our final expression for the amplitude as a function on the
decorated moduli space (note that it is only non-zero in the region where
$\Delta > 0$).
The generalization to the contribution of this diagram to $\langle
\tr(\Phi^{J_1}(x_1)) \tr(\Phi^{J_2}(x_2)) \tr(\Phi^{J_3}(x_3))\rangle_{S^2}$ is
simply given by adding appropriate powers of $\sigma_i$'s in (\ref{posDelta}).
%
%
 \item $\Delta=0$
%
%
: The topology of the corresponding gauge theory graph in this
case is given by Figure \ref{regs3PointF}(b). This graph
contributes, for instance, to correlation functions of the form
$\vev{\tr(\Phi^{J_1}(x_1)) \tr(\Phi^{J_1+J_2}(x_2))
\tr(\Phi^{J_2}(x_3))}_{S^2}$, for which it is the only
contributing graph. For example, the $\vev{\tr(\Phi^{2}(x_1))
\tr(\Phi^{4}(x_2)) \tr(\Phi^{2}(x_3))}_{S^2}$ correlation function
is given by
 \be
 G &=&
{\tilde A} \int d\sigma_1 d\sigma_2 \sigma_1 \sigma_2
 e^{-\frac{(x_2-x_1)^2\sigma_1}{4}}e^{-\frac{(x_2-x_3)^2\sigma_2}{4}}
\nonumber\\
 &=&
{\tilde A} \int dp_0 dp_1 p_0 p_1
 e^{-\frac{(x_2-x_1)^2p_0}{4}}e^{-\frac{(x_2-x_3)^2p_1}{4}}
=
\frac{16 {\tilde A}}{(x_2-x_1)^4(x_3-x_2)^4}
 \ee
for some constant ${\tilde A}$. Again,
the generalization to more general correlation functions is straightforward.
%
%
%
Note that the non-zero contribution of this diagram comes just from the
two-dimensional subspace of the decorated moduli space with $p_{\infty}
=p_0+p_1$.
 \item $\Delta<0$
: The topology of the corresponding graph is given by Figure
\ref{regs3PointF}(c).  It has a (topologically non-trivial)
self-contraction, so when all operators are normal-ordered it
vanishes. Thus, all diagrams vanish in this part of the decorated
moduli space.
 \end{itemize}

To summarize,
when we calculate an amplitude
$\langle \tr(\Phi^{J_1}(x_1)) \tr(\Phi^{J_2}(x_2)) \tr(\Phi^{J_3}(x_3))
\rangle_{S^2}$
with no self-contractions, the multiplicities $m_{ij}$ of the edges are given
by
\be m_{ij}=\half(\sum_kJ_k)-\sum_k|\epsilon_{ijk}|J_k. \ee
We always get a contribution from only one region of the decorated moduli
space.
If one of the $m_{ij}$ is zero we find a
contribution only from $\Delta=0$, while if all the
$m_{ij}>0$ the only contribution comes from $\Delta>0$.
In other cases the amplitude vanishes.
%
%
\subsection{The $Y$ four-point function diagram}
Now that we understand how the formalism works on the simple
three-point function example we turn to four-point functions.  Our
main interest will be in studying the worldsheet OPE. Studying this
OPE using four-point functions is somewhat subtle, since in a two
dimensional conformal field theory, the non-trivial dependence on the
positions in four-point functions is a function only of
the cross-ratio $z=(z_1-z_2)(z_3-z_4)/(z_1-z_3)(z_2-z_4)$, and the
$z\to 0$ limit can be interpreted either as $z_1\to z_2$ or as $z_3\to
z_4$. We will see that one of the interpretations will turn out to be
more natural in the computation of this subsection.

Finding the Strebel differential for a general four-point diagram
is a difficult task. The most general Strebel differential for a
four-point amplitude, when we choose the insertions to be at
$z=\pm t, 1, \infty$, has the following form \cite{Gopakumars}:
\be \label{gen4point}
q=-\frac{p_\infty^2}{4\pi^2}\frac{\prod_{i=1}^4(c_i-z)}{(1-z)^2(t^2-z^2)^2}
dz^2.  \ee The sphere four-point function has a single complex
modulus $t$, and there are four circumferences $p_i$. The
worldsheet OPE appears in this parametrization as $t\to 0$. Recall
that the field theory amplitude is given in terms of the edge
lengths of the edges connecting the zeros. These are integrals of
the square root of (\ref{gen4point}).  The differential is Strebel
when these edge lengths are real, and this gives very non-trivial
equations on the parameters of (\ref{gen4point}). In the rest of
this section
we will consider special cases where the Strebel problem can be
explicitly solved.
\begin{figure}[tbp]
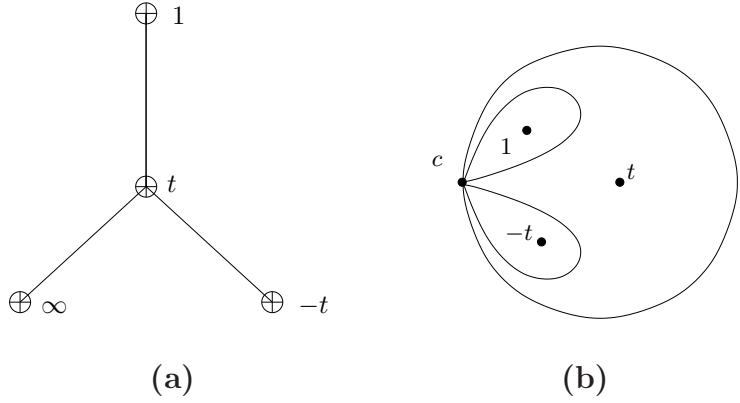

\begin{center}
$\begin{array}{c@{\hspace{.5in}}c}
\input{Ydiag2.pstex_t} & \input{YdiagDual2.pstex_t}
 \\ [0.2cm]
\mbox{\bf (a)} & \mbox{\bf (b)}
\end{array}$
\caption{The $Y$ diagram  with $t$ in
the middle:
{\bf (a)} in the gauge theory, {\bf (b)} on the worldsheet.} \label{Y2}
\end{center}
\end{figure}

The first case we consider is the $Y$ diagram of the gauge theory,
drawn in Figure \ref{Y2}(a). We arbitrarily mapped the operator in
the middle to the point $t$ on the worldsheet, meaning that we
compute the OPE between the operator in the middle and another on
an edge; as discussed above, we cannot distinguish this from an
OPE between the two other operators (which are both on edges) so
this is in fact the most general possibility. The gauge theory
scalar correlators which get contributions from this diagram are:
\be \label{Yfieldtwo} \langle \tr(\Phi^{m_t}(0))
\tr(\Phi^{m_{\infty}}(x_{\infty})) \tr(\Phi^{m_{-t}}(x_{-t}))
\tr(\Phi^{m_1}(x_1))\rangle_{S^2} \ee (putting $x_t=0$ for
convenience), with $m_t = m_{\infty} + m_{-t} + m_1$,
and this is the only diagram contributing to these correlators.
The dual diagram, drawn (schematically)
in Figure \ref{Y2}(b), consists of three disjoint
disks glued at a point. Since six horizontal lines join at the vertex,
it is a degenerate zero of the Strebel differential of order four.
Thus, the Strebel differential for this diagram is given by:
\begin{equation}\label{differential 3 edges of the form Y}
    q=-\frac{p_{\infty}^2}{4\pi^2}\left(\frac{(z-c)^2}{(z-1)(z^2-t^2)}\right)^2
dz^2.
\end{equation}
We have three edges in this diagram, so the diagram will be non-zero
only on a three real-parameter subspace of the six-real-dimensional
decorated moduli space. From the figure it is clear that the four
circumferences satisfy one constraint among themselves,
\begin{equation}
\label{circrel}
    p_t=p_{\infty}+p_1+p_{-t},
\end{equation}
so only three are independent; we can thus label the position in the subspace
where the diagram is non-zero either by three of the circumferences, or
by $t$ and one of the three circumferences.
In general to determine the $p_i$ from the Strebel differential we
need to take a square root of the residues $p_i^2$, and there is a
sign ambiguity. However, in this case (\ref{circrel})
fixes the signs of the square roots uniquely :
\be
\frac{p_t}{p_{\infty}}=-\frac{(c-t)^2}{2t(-t+1)},\qquad \frac
{p_1}{p_\infty}=-\frac{(c-1)^2}{1-t^2},\qquad
\frac{p_{-t}}{p_\infty}=-\frac{(c+t)^2}{2t(t+1)}.
\ee
Defining
\be
A=\frac{p_{-t}}{p_{\infty}}-\frac{p_{ t}}{p_{\infty}},\qquad
B=\frac{p_{t}}{p_{\infty}}+\frac{p_{-t}}{p_{\infty}},
\ee
we have
\be
    c=\half\left(-A-tB\right),\qquad
    t^2\left(4+B^2+4A\right)+2t\left(2B+AB\right)+A^2=0.
\ee
The discriminant of this quadratic equation with real coefficients,
$\Delta=16(B^2-A^2)(A+1)$, is
always non-positive because
of the obvious inequalities $|B|\geq |A|$ and $A\leq -1$, so the solutions
for $t$ as a function of the edge-lengths are generally complex (as expected).

Next, we would like to work out the change of variables from the
Schwinger parameters (which are simply related to the edge
lengths, which for this diagram are equal to $p_1$, $p_{\infty}$
and $p_{-t}$) to $t$ and $p_{\infty}$, in order to be able to
study the OPE limit $t\to 0$. The change of variables to
$p_{\infty}$, $A$ and $B$ is straightforward.
Next, we can translate $A$ and $B$ to $t$ by using\footnote{It is
important to emphasize that two given metric graphs of the form
above which differ only in the cyclic order of the points around
the graph are not in the same isomorphism class, and hence they
correspond to two distinct Riemann surfaces. These two Riemann
surfaces are related by complex conjugation of $t$.} \be {\rm
Re(t)}=-\frac{2B+BA}{4+B^2+4A},\qquad |t|^2=\frac{A^2}{4+B^2+4A}.
\ee
With some algebra
we find that the measure on the decorated moduli space transforms as
\begin{equation}
\label{measure}
\int dp_\infty dAdB=\int dp_\infty
d^2 t |{\rm Im}(t)|
\left|\frac{4A((A+2)^2-B^2)}{(B^2+4A+4)^3}\right|^{-1}.
\end{equation}

Next, we examine the field theory amplitude, and integrate over
$p_\infty$. The field theory amplitude (\ref{Yfieldtwo}) in
position space, when expressed in terms of the inverse Schwinger
parameters, is given up to a constant by
\begin{gather}\label{leafmiddlemomentum}
    G=\int dp_\infty dp_{-t} dp_1
    p_\infty^{(m_\infty-1)} p_{-t}^{(m_{-t}-1)} p_1^{(m_1-1)}
    e^{-p_\infty(x_{\infty}^2
+x_{-t}^2(B+A)/2-x_1^2(1+A) )},
\end{gather}
which may be rewritten (after changing variables using
(\ref{measure}) and integrating over
$p_{\infty}$) in the convenient form
\begin{gather}\label{leafmiddlespace}
    G=\int
d^2 t |{\rm Im}(t)|
    (x_{\infty}^2
+x_{-t}^2(B+A)/2-x_1^2(1+A)
)^{-m_\infty-m_1-m_{-t}}\times\cr\times
\left|\frac{4A((A+2)^2-B^2)}{(B^2+4A+4)^3}\right|^{-1}
(A+B)^{(m_{-t}-1)}
    (1+A)^{(m_1-1)}.
    \end{gather}

We can now consider the OPE limit $t\to 0$. Writing
$t=|t|e^{i\phi}$, we find that for small $t$ we have
\begin{equation}\label{phase leafmiddle}
    B=\frac{2}{(1- \cos(\phi))|t|}+O(1), \hspace{2em}
    A=\frac{-2}{1-
    \cos(\phi)}+O(|t|).
\end{equation}
%
Substituting this into
(\ref{leafmiddlespace}) we find that the
leading term in the OPE which contributes to this diagram
is given by (up to a constant) :
\begin{equation}
\label{phaseleaves}
\int d^2t
|t|^{m_t-m_{-t}-2}|\sin(\phi)|(1+\cos(\phi))^{m_1-1}
(1-\cos(\phi))^{m_\infty-1}(x_{-t}^2)^{-m_t}.
\end{equation}
As expected, only powers of $|t|$ bigger than $(-2)$ appear, since the
integration over $t$ must be convergent. From the form of
(\ref{phaseleaves}) we can read off
the worldsheet conformal dimensions $(h, \bar h)$ of the operators
appearing in the leading term: generically we find operators with
\be
h+\bar h=m_t-m_{-t}+2, \qquad h=\frac{q}{2}\ (q\in \mathbb{Z}),\qquad
h,\bar h\geq \frac{3}{2}.
\ee
%
For instance, for $m_t=7$, $m_{-t}=2$, we have
\be (h,\bar h)=
(\frac{3}{2},\frac{11}{2}),(2,5),(\frac{5}{2},\frac{9}{2}),(3,4),
(\frac{7}{2},\frac{7}{2}),(4,3),(\frac{9}{2},\frac{5}{2}),(5,2),
(\frac{11}{2},\frac{3}{2}).
\ee

Note that, as expected, the OPE coefficient becomes singular as
$x_t$ approaches $x_{-t}$ (this is why it is more natural to
interpret this OPE in terms of these two points approaching each
other, rather than the other two points). As discussed in section
\ref{OPEsect}, when the space-time theory has a conformal symmetry
(as in our case), we would expect that for every value of $t$ the
worldsheet correlation function should transform under the
conformal symmetry in the same way as the full space-time
correlation function (given by the integral over $t$). However, it
is easy to see that this is not the case; the ratio between the
expression $(x_{-t}^2)^{-m_t}$ which we found for small $t$ and
the exact field theory answer $(x_{-t}^2)^{-m_{-t}}
(x_{\infty}^2)^{-m_{\infty}} (x_1^2)^{-m_1}$ cannot be written as
a function of the two cross-ratios $(x_{\infty}-x_{-t})^2 x_1^2 /
(x_{\infty}-x_1)^2 x_{-t}^2$ and $x_{\infty}^2 (x_{-t}-x_1)^2 /
(x_{\infty}-x_1)^2 x_{-t}^2$, as conformal invariance would
demand. However, our small $t$ expression does have the correct
scaling transformation (consistent with our discussion in section
\ref{OPEsect}).

Next, we calculate the subleading term in the expansion of the
OPE.
The leading correction to (\ref{phase leafmiddle}) is given by
\begin{align}
    B&=\frac{2}{(1- \cos(\phi))|t|}+\frac{2|t|(\cos(\phi)-\half)(\cos(\phi)+1)}
{\cos(\phi)-1} +O(|t|^2), \cr
    A&=\frac{-2}{1-\cos(\phi)}-\frac{|t|^2\cos(\phi)(\cos(\phi)+1)}
{\cos(\phi)-1}+O(|t|^3).
\end{align}
Using (\ref{leafmiddlespace}) we find that
the subleading term in
the worldsheet OPE  is
\begin{gather}
\int d^2t
|t|^{m_t-m_{-t}-1}|\sin(\phi)|(1+\cos(\phi))^{m_1-1}(1-\cos(\phi))^{m_\infty-1}
(x_{-t}^2)^{-m_t-1}
\times\cr\times\left(
x_{-t}^2(1-m_\infty-m_1)+x_\infty^2(\cos(\phi)-1)m_t
-x_1^2(\cos(\phi)+1)m_t\right).
\end{gather}
The qualitative behavior is similar to that of the leading term,
but now we find a different space-time dependence for different
terms in the OPE (with the same $h+{\bar h}$ but different $h$),
as we would generally expect.

Even though this diagram is quite simple, we showed that it is
possible to use it to study many of the features of the
gauge-string duality that we are trying to understand.

\subsection{The $\Pi$ four-point function diagram}

\begin{figure}[tbp]
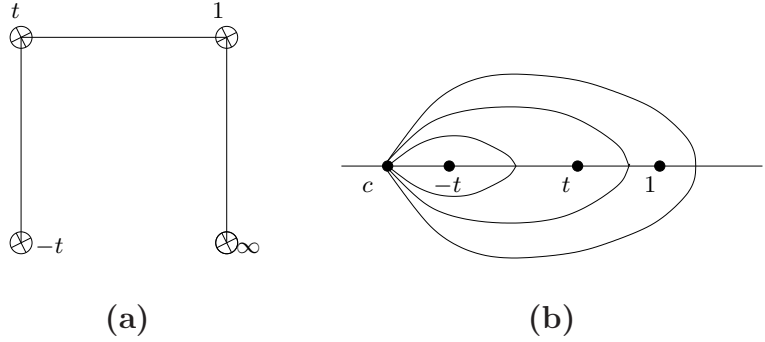

\begin{center}
$\begin{array}{c@{\hspace{.5in}}c}
\input{Pigraph.pstex_t} & \input{Picrit3.pstex_t}
 \\ [0.2cm]
\mbox{\bf (a)} & \mbox{\bf (b)}
\end{array}$
\caption{The $\Pi$ diagram:
{\bf (a)} in the gauge theory, {\bf (b)} on the worldsheet.} \label{Pi}
\end{center}
\end{figure}

The next solvable four-point function we consider is the other
diagram with three edges, which we call the $\Pi$ diagram (Figure
\ref{Pi}). It contributes  to various correlators in general field
theories; for example, in a free gauge theory with three adjoint
scalar fields $\Phi_1,\Phi_2,\Phi_3$ it is the only contribution
to the correlator : \be \label{Pifield} \langle \tr(\Phi_1^2(x_1))
\tr(\Phi_1^2\Phi_2(x_2)) \tr(\Phi_2\Phi_3^2(x_3))
\tr(\Phi_3^2(x_4))\rangle_{S^2}. \ee We will see that this diagram
has a very strange property --
the string
theory amplitude has support only for real values of $t$
(namely, when all four points lie on a straight line in the plane).

The Strebel differential here is again given by:
\begin{equation}\label{differential 3 edges}
    q=-\frac{p_{\infty}^2}{4\pi^2}\left(\frac{(z-c)^2}{(z-1)(z^2-t^2)}\right)^2
dz^2,
\end{equation}
but with a different relation between the circumferences.
When we label the points as in Figure \ref{Pi}, we have
\be
\label{choice}
p_{t}-p_{-t}=p_1-p_{\infty}.
\ee
The only possible sign choice is then
\be
\frac{p_{\pm t}}{p_{\infty}}=\frac{(c\mp t)^2}{2t(\mp t+1)},\qquad
\frac{p_1}{p_\infty}=\frac{(c-1)^2}{1-t^2}.
\ee
Again, it is convenient to define
\be
A=\frac{p_{-t}}{p_{\infty}}-\frac{p_{ t}}{p_{\infty}},\qquad
B=\frac{p_{t}}{p_{\infty}}+\frac{p_{-t}}{p_{\infty}}.
\ee
Then, we have
\be
\label{eqfort}
2c=A+tB,\qquad (A+tB)^2=4t(B+tA-t).
\ee
The last equation gives $t$ for a given value of the circumferences.
We would naively expect to
have solutions for all values of $t$ near zero, in order to
get an expression with a smooth OPE expansion.
However, this turns out not to be the case; equation (\ref{eqfort}) gives :
\begin{equation}\label{quadequation}
t^2(B^2-4A+4)+2t(AB-2B)+A^2=0.
\end{equation}
This is a quadratic equation with real coefficients.
The discriminant is
\begin{equation}
\Delta=4(AB-2B)^2-4A^2(B^2-4A+4)=
16(A^2-B^2)(A-1).
\end{equation}
This is always
non-negative by the obvious inequalities $B\geq |A|$ and $A\leq 1$. So, these
diagrams always correspond to real values of $t$, covering just a
one-dimensional subspace of the moduli space.
Thus, in this case, if we translate a correlation function such as
(\ref{Pifield}) to the worldsheet, we find a non-smooth correlation
function on the worldsheet which is non-zero only on a one-dimensional
subspace. Obviously, this does not have a good OPE expansion, and it
cannot arise from a sensible local field theory on the worldsheet.
Perhaps there are some global zero modes causing this worldsheet
correlation function to vanish for generic values of $t$.
\begin{figure}[htbp]
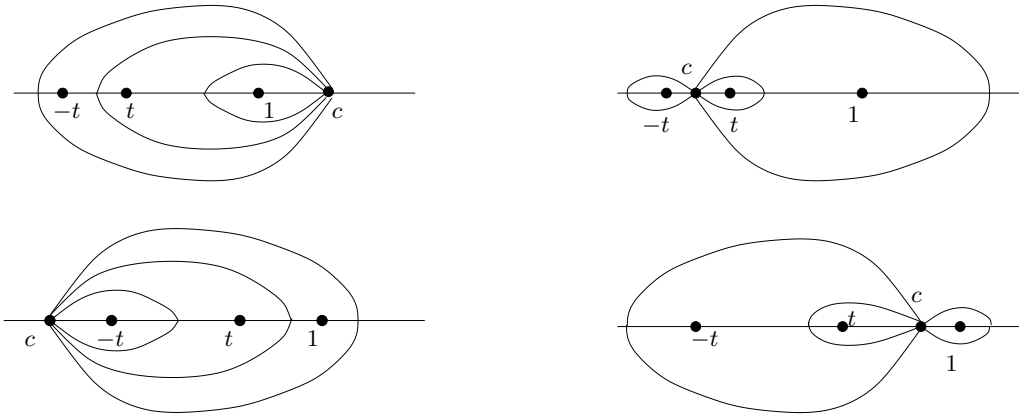

\begin{center}
 $\begin{array}{c@{\hspace{1in}}c}
\input{Picrit1.pstex_t} &\input{Picrit2.pstex_t}
 \\ [0.4cm]
\input{Picrit3.pstex_t} &\input{Picrit4.pstex_t}
 \\ [0.4cm]
\end{array}$
\caption{The four possible positions of the zero $c$ and the
structure of critical leaves implied by the symmetry of
reflections round the real axis. The two critical curves on the
left (and on the right)
 are different as metric graphs because they have a different orientation.}
\label{Picrit}
\end{center}
\end{figure}

\begin{figure}[tbp]
\begin{center}
\input{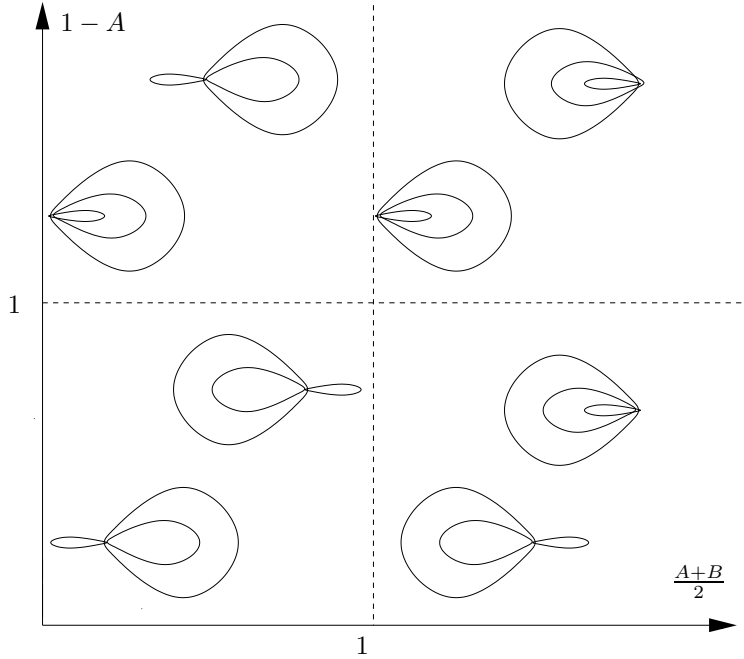}
\caption{The four regions of $A$,$B$ space;
for each $A$,$B$ there are two possible solutions of the quadratic
equation, and we draw the two distinct diagrams that they corresponds to. As
explained, each metric graph (including orientation)
appears exactly once.} \label{modulispace4point}
\end{center}
\end{figure}

Note that our choice (\ref{choice}) by itself does not uniquely
determine the topology of the critical curve; there are four
topologically distinct critical curves obeying this relation.
We draw these four possibilities in Figure \ref{Picrit}.
The differential in this case is real, so
the critical lines are obviously invariant under reflection with respect to
the real axis.
At first sight one may worry that we obtain from the quadratic equation
(\ref{quadequation}) above two different
solutions for $t$ with the same circumferences, even though the
Strebel isomorphism should give a unique (marked) Riemann surface
for every value of the edge lengths. The resolution is that the
two solutions have different topologies, so they correspond to
different graphs (Feynman diagrams) on the field theory side.
The summary of the
possible graphs which appear for each value of the circumferences
(labeled by $A$ and $B$) is given in Figure
\ref{modulispace4point}.

\subsection{The $X$ five-point function diagram}\label{SphereX}

The final example we discuss in detail is the five-point $X$
amplitude, drawn in Figure \ref{star}. This is the single
contribution to correlators of the form \be \label{Xcorrelator}
\langle \tr(\Phi^{m_1}(x_1))\tr(\Phi^{m_b}(x_b))
\tr(\Phi^{m_{\infty}}(x_{\infty}))\tr(\Phi^{m_{-t}}(x_{-t}))\tr(\Phi^{m_t}
(x_t))\rangle_{S^2}, \ee with $m_t=m_1+m_b+m_{\infty}+m_{-t}$.
Here we will illustrate how the fact that the string theory
amplitude localizes on a sub-manifold of the moduli space makes it
difficult to extract the OPE.
\begin{figure}[htbp]
\begin{center}
$\begin{array}{c@{\hspace{.5in}}c}
\input{star.pstex_t} & \input{XdiagDual.pstex_t}
 \\ [0.2cm]
\mbox{\bf (a)} & \mbox{\bf (b)}
\end{array}$
\caption{The $X$ diagram:
{\bf (a)} in the gauge theory, {\bf (b)} on the worldsheet.} \label{star}
\end{center}
\end{figure}

The Strebel differential in this case has a zero of order six, so
it is given by (choosing the insertions at $z=1,\pm t, b,
\infty$): \be
q=-\frac{p_\infty^2}{4\pi^2}\frac{(c-z)^6}{(1-z)^2(b-z)^2(t^2-z^2)^2}
dz^2. \ee
The ratios of the circumferences are given by (denoting
$p_t=p_+$, $p_{-t}=p_-$)
\be
\frac{p_1}{p_\infty}&=&\gamma_1 \frac{(c-1)^3}{(1-t^2)(b-1)}\equiv \a_1^3,\quad \frac{p_b}{p_\infty}=\gamma_b \frac{(c-b)^3}{(b^2-t^2)(b-1)}\equiv \a_b^3,\nonumber\\
\frac{p_+}{p_\infty}&=&\gamma_+ \frac{(c-t)^3}{2t(1-t)(b-t)}\equiv
\a_+^3,\quad \frac{p_-}{p_\infty}=\gamma_-
\frac{(c+t)^3}{2t(1+t)(b+t)}\equiv \a_-^3,
\ee
up to signs $\gamma_i$, which depend on the choice of which vertex
we put in the middle; note the useful identity
\be \frac{(c-1)^3}{(1-t^2)(b-1)}  -\frac{(c-b)^3}{(b^2-t^2)(b-1)}
- \frac{(c-t)^3}{2t(1-t)(b-t)} + \frac{(c+t)^3}{2t(1+t)(b+t)}-1=0.
\ee
%
If we put the vertex at the point $j$
in the middle we have:
\be
\sum_{i\neq j}\frac{p_i}{p_j}-1=0.
\ee
There are two possible cases to consider when taking the small $t$
limit : either putting one of $\pm t$ in the middle, or putting both
on the edges.

On the field theory side we have four edge-length parameters,
which should map to a subspace of the decorated moduli space (the
circumferences, $b$ and $t$). As discussed in section
\ref{Preliminaries} we have a $p_i \to \alpha p_i$ scaling
symmetry which does not change the Riemann surface (and rescales
the edge lengths), so we will cover at most a three
real-dimensional subspace of the moduli space (the space of $b$'s
and $t$'s).

As in the previous examples, we can solve the equations above to
get all the constraints, without any need to invoke the reality
conditions of the Strebel differential (which are automatically
satisfied). We define
\be F_1=\gamma_1^{1/3}\biggl((1-t^2)(b-1)\biggr)^{1/3},\quad
F_\pm=\gamma_\pm^{1/3}\biggl(2t(1\mp t)(b\mp t)\biggr)^{1/3}, \ee
where we choose a specific root for the third root of the expressions
in the brackets,
and the ambiguity is
swallowed in the third root of the $\gamma$'s. The equations for the
circumferences become:
\be \a_1F_1+1=c,\quad \a_+F_++t=c,\quad \a_-F_--t=c. \ee
We take
all the $\a_i$ to be real and positive. The first equation can be
taken to be an equation for $c$, and from the two others we get four
real equations. Two are equations for $\a_\pm$ :
\be
2t_1=\a_-{\rm Re}(F_-)-\a_+{\rm Re}(F_+),\quad
2t_2=\a_-{\rm Im}(F_-)-\a_+{\rm Im}(F_+),
\ee
where
$t_1\equiv {\rm Re}(t)$ and  $t_2\equiv {\rm Im}(t)$.
One equation gives $\a_1$,
\be
\a_1={\rm Re}\biggl(\frac{t+\a_+F_+-1}{F_1}\biggr), \ee
and the last one
is a constraint on the moduli space:
\be\label{constraint 5 point}
0={\rm Im}\biggl(\frac{t+\a_+F_+-1}{F_1}\biggr). \ee
The equations for $\a_\pm$ are easily solved:
\be
\a_\pm=2\frac{t_2{\rm Re}(F_\mp)-t_1{\rm Im}(F_\mp)}{{\rm Re}(F_+)
{\rm Im}(F_-)-{\rm Re}(F_-){\rm Im}(F_+)}.
\ee

We begin with the case when the two $\pm t$ insertions are on the
edges. Without loss of generality we take $\infty$ to be in the
middle. For this choice the signs are \be
\gamma_1=-\gamma_b=-\gamma_+=\gamma_-=1. \ee To the leading order
in $t$ we then get :
\be F_\pm=(2|tb|)^{1/3}e^{\frac{i}{3}(\pi k_\pm+\phi+\theta_b)},
\ee
where $k_+$ is an odd integer and $k_-$ is even, $\phi$ is the phase
of $t$, and $\theta_b$ is the
phase of $b$. From here we find
\be
\Delta&\equiv&{\rm Re}(F_+){\rm Im}(F_-)-{\rm Re}(F_-){\rm Im}(F_+)=\nonumber\\
&=&(2|tb|)^{2/3}\biggl(\cos(\frac{\pi
k_++\phi+\theta_b}{3})\sin(\frac{\pi k_-+\phi+\theta_b}{3})
-\sin(\frac{\pi k_++\phi+\theta_b}{3})\cos(\frac{\pi k_-+\phi+\theta_b}{3})\biggr)\nonumber\\
&=&(2|tb|)^{2/3}\sin\biggl(\frac{\pi (k_--k_+)}{3}\biggr).
\ee
Note that if $k_--k_+=3m$ we get a contradiction : we find
that $p_\pm$ should go as $\pm B/|t|$ for small $t$, which contradicts the
positivity of the circumferences (because the zero $c$ goes to a
constant as $t\to 0$ if $k_--k_+=3m$). Thus, the $\a_\pm$ are:
\be \a_\pm&=&\frac{2(2|t|)^{2/3}}{|b|^{1/3}\sin\biggl(\frac{\pi
(k_--k_+)}{3}\biggr)}
\biggl(\sin(\phi)\cos(\frac{\pi k_\mp+\phi+\theta_b}{3})-\cos(\phi)\sin(\frac{\pi k_\mp+\phi+\theta_b}{3})\biggr)\nonumber\\
&=&2(2|t|)^{2/3}\frac{\sin\biggl(\frac{1}{3}(2\phi-\pi
k_\mp-\theta_b)\biggr)}{|b|^{1/3}\sin\biggl(\frac{\pi
(k_--k_+)}{3}\biggr)}. \ee
The circumferences are: \be
\frac{p_\pm}{p_\infty}=\frac{8(2|t|)^2}{|b|}\biggl(\frac{\sin\biggl(\frac{2\phi-\pi
k_\mp-\theta_b}{3}\biggr)}{\sin\biggl(\frac{\pi
(k_--k_+)}{3}\biggr)}\biggr)^3. \ee
The phases $k_\pm$ are set
unambiguously by demanding the positivity of these expressions. The
phase choices are:
\be
&&\phi-\frac{\theta_b}{2}\in(0,\pi) \to (k_-,k_+)=(0,5),(2,3);\\
&&\phi-\frac{\theta_b}{2}\in(\pi,2\pi) \to
(k_-,k_+)=(4,5),(2,1).\nonumber \ee
There are four choices of the phases which give exactly the same
results. The different choices are simply switching $p_+$ and
$p_-$ (relabeling the edges).
We note that here $c\to 0$ as $t\to 0$, and thus
$b=-p_b/p_1\leq 0$. Thus, $b$ is real and negative.
The Jacobian for the change of variables from $p_+/p_{\infty},
p_-/p_{\infty}$ to $|t|,\phi$ is easily obtained from here : for small
$|t|$
%
\be {\mathcal J}\sim |t|^3\frac{\sin\biggl(\frac{1}{3}(2\phi-\pi
k_+)\biggr)^2\sin\biggl(\frac{1}{3}(2\phi-\pi
k_-)\biggr)^2}{\sin\biggl(\frac{\pi (k_--k_+)}{3}\biggr)^5}. \ee
Note also that the constraint (\ref{constraint 5 point})
on the moduli space reads
\be
0={\rm Im}\biggl[\frac{1}{F_1}\biggr]\to{\rm Im}
\biggl[\gamma_1^{1/3}(1-b)\biggr]=0,
\ee
so we get
that  $\gamma_1^{1/3}=1$. We also have two positivity constraints
for $p_1$, $p_b$ which give
\be
-1\leq{\rm Re}\biggl[\frac{1}{F_1}\biggr]\leq 0
\to -1\leq\frac{1}{(b-1)^{1/3}}\leq 0,
\ee
which is satisfied. Thus all the phases are set.

Let us now discuss the case of one $\pm t$ insertion in the
middle, say $t$. We obtain:
\be F_\pm=e^{\frac{\pi i}{3}k_\pm}(2tb)^{1/3}(1\mp
(1+b^{-1})\frac{1}{3}t). \ee
In this case
\be
\gamma_1=-\gamma_b=\gamma_{t}=\gamma_{-t}=-1.
\ee
Note that if $k_+\neq k_-$ the equations are essentially the same
as above. Thus, the $p_\pm$ circumferences go to zero as $t$ goes
to zero. This, however, implies that the other circumferences
cannot all be positive because we have
$p_+=p_-+p_1+p_b+p_\infty$. Thus, this case is inconsistent and we
must have here $k_\pm=k$ where $k$ is odd or even depending on which
insertion, $\pm t$, we chose to put in the middle. Thus we find
(at leading order in $|t|$)
\begin{equation}
\a_\pm=\frac{3}{(2|tb|)^{1/3}}\biggl(\frac{\sin\biggl(\frac{1}{3}(2\phi-\pi
k-
\theta_b)\biggr)}{|1+b^{-1}|\sin\biggl(\phi+\tilde\theta_b\biggr)}\pm
\frac{|1+b^{-1}|\sin\biggl(\frac{1}{3}(\phi+\pi
k)+\theta_b+\tilde\theta_b\biggr)}{\sin\biggl(\phi+
\tilde\theta_b\biggr)}|t|\biggr),
\end{equation}
where $\tilde\theta_b$ is the phase of $(1+b^{-1})$ and
$\theta_b$ is the phase of $b$.
The circumferences
behave as:
\begin{equation}
\frac{p_\pm}{p_\infty}=
\frac{27}{2|tb|}\biggl(\frac{\sin\biggl(\frac{1}{3}(2\phi-\pi
k-\theta_b)\biggr)}{|1+b^{-1}|\sin\biggl(\phi+\tilde\theta_b\biggr)}\biggr)^3\biggl(1\pm
3|1+b^{-1}|^2\frac{\sin\biggl(\frac{1}{3}(\phi+\pi
k)+\theta_b+\tilde\theta_b\biggr)}{\sin\biggl(\frac{1}{3}(2\phi-\pi
k-\theta_b)\biggr)}|t|\biggr). \end{equation}
Note that $k$ should be chosen such that the expression will be
positive, and it depends on $\tilde\theta_b$. The details are similar to
the previous case but the expressions are more complicated so we omit them
here. There is only one choice of $k$ such that the solution is
well-behaved as a function of $\phi$. Also note that here
$\a_1$ depends on the phase $\phi$. The Jacobian is a
complicated function of the phases, because
even the leading term of $\a_1$ has a non-trivial dependence on
$\phi$, and $b$ is also a complicated function of $\phi$.
At leading order as $t\to 0$ we find
\be
{\mathcal
J}=J_0(\theta_b,|b|,\phi)\frac{1}{|t|^2},
\ee
where the Jacobian is
from the integration over $p_i/p_\infty$ to the integration over
$d^2bd|t|d\phi$.

We now calculate the string theory amplitude from the gauge theory
expression as usual, for the correlation function
(\ref{Xcorrelator}).
In the gauge theory we have (up to a constant)
\be
\label{Xresult}
G&=&\int\prod_{i=-t,b,1,\infty}d\s_i\s_i^{m_i-1}e^{-\frac{(x_i-x_t)^2}{4}\s_i}
\nonumber\\
&=&\int\prod_{i=-t,b,1}
d\frac{p_i}{p_\infty}(\frac{p_i}{p_\infty})^{m_i-1}\int dp_\infty
p_\infty^{\sum_{i=-t,b,1,\infty}(m_i-1)+3}
e^{-p_\infty A}\nonumber\\
&\propto&\int\prod_{i=-t,b,1}
d\frac{p_i}{p_\infty}(\frac{p_i}{p_\infty})^{m_i-1}\biggl(\frac{1}{A}
\biggr)^{m_t},
\ee
where
\be
A\equiv \sum_{i=-t,b,1,\infty} \frac{(x_i-x_{t})^2}{4}\frac{p_i}{p_\infty}.
\ee
We find that in the $t\to 0$ limit the amplitude has the simple form :
\be
G&\propto&\int d^2 t\frac{1}{(x_t-x_{-t})^{2m_t}}
|t|^{m_1+m_b+m_\infty-2}\tilde F(\phi)\\
&=&\int d^2 t\frac{1}{(x_t-x_{-t})^{2 m_t}}|t|^{m_t-m_{-t}-2}\tilde
F(\phi).\nonumber
\ee
The dependence on the phases here is quite complicated. The
leading divergent power in the OPE
agrees with what we found for the $4$-point $Y$ amplitude (\ref{phaseleaves}).
Note that, unlike what we found before,
the dependence on the phase $\phi$ is not a finite sum of $\sin$'s.
The easy way to see
this is to note that the only dependence on $m_\infty$ in (\ref{Xresult})
comes from
the term
\be
\biggl(\frac{1}{A}\biggr)^{m_\infty}\sim\biggl(\frac{1}{(x_t-x_{-t})^{2}
\frac{p_{-t}}{p_\infty}}\biggr)^{m_\infty}
\ee
in
the integrand. From the calculation of $\frac{p_{-t}}{p_\infty}$
above we know that the phase dependence of this term is of
the form
\be
\biggl(\frac{\sin\biggl(\phi+\tilde\theta_b\biggr)}{\sin
\biggl(\frac{1}{3}(2\phi-\pi
k-\theta_b)\biggr)}\biggr)^{3m_\infty}.
\ee
This expression
does not factorize as a finite sum of trigonometric functions.
However, it turns out that
this is a meaningless statement because we have the
delta function constraint
(\ref{constraint
5 point})
(recall that the integral is only over a
three-dimensional subspace of the space of $b$'s and $t$'s),
which is a functional relation among $\phi$, $\theta_b$
and $|b|$ (to the leading order in $|t|$), and thus, we can substitute
anything which equals one on the constraint space into the integrand. The
result is that the $\phi$ dependence is not well defined here.
On the other hand, the power of $|t|$ appearing in the OPE
is well-defined,
because the delta function for small $|t|$ is independent of
$|t|$. This is not trivial; if the constraint had looked like (for
small $|t|$) $\delta(\frac{\cdot\cdot\cdot}{|t|})$, then the power
of $|t|$ in the most singular term would have also been an ill-defined
quantity.

An interesting point to note is that for the previous case, with
the $\pm t$ insertions on the edges, the series begins with more
regular terms than in the other OPEs we analyzed so far. We find
that the most singular term in this case is given by :
\be
\label{subleadingX}
G\sim\int d^2 t|t|^{2(m_t+m_{-t})-2}\tilde F_2(\phi,x_i).
\ee
Our interpretation of this is that the leading terms in the OPE, going
as $|t|^{|m_t-m_{-t}|-2}$ and higher powers, do not contribute to the
correlation function in this case, and the terms which contribute start
at the order appearing in (\ref{subleadingX}).

It is easy to generalize the procedure of this subsection to
general $star$ diagrams, with one point in the middle connected to
$n-1$ other points by edges. For an $n$-point $star$ diagram, if
we put $\pm t$ on the edges, we find \be
\frac{p_\pm}{p_\infty}\sim |t|^{n-3}, \quad{\mathcal J}\sim
|t|^{2(n-4)}, \ee and we will get even higher powers of $|t|$ as
the first terms of the OPE expansion. It is still true, however,
that the power will depend only on the combination $m_t+m_{-t}$.
On the other hand, if we put $t$ in the middle and $-t$ on the
edge, we reproduce the same $|t|$ power that we found (for
$n=4,5$) in the $Y$ and $X$ diagrams. We find here:
\be\frac{p_\pm}{p_\infty}\sim\frac{1}{|t|},\quad{\mathcal J}\sim
|t|^{-2},\ee leading to a leading power of $t$ in the OPE equal to
$m_t-m_{-t}-2$.

\section{Two-Point Function on the Torus}
In this section we will
discuss some aspects of the two-point function on the torus (note
that while in a general string theory one could have also non-zero
one-point functions on the torus, these are not present in free
gauge theories with normal-ordered operators so we ignore them
here).  In string theory, a general two-point function on a torus
has four real parameters, the complex position $b$ of one of the
insertions (when the other insertion is chosen at $z=0$) and the
complex torus modulus $\tau$. In the decorated moduli space there
are two additional moduli, the two circumferences related to the
two insertions. Thus, we have six real parameters for a two-point
function on the torus.

If we map a generic point on this decorated moduli space to a
field theory two-point function using the general prescription
described in section \ref{Preliminaries}, we find a toroidal
Feynman diagram with six edges, but with two (topologically
non-trivial) self-contractions.
A toroidal 2-point function with no self-contractions has only
four edges in its ``skeleton graph'' (see, for instance,
\cite{Constable:2002hw} for
discussions of toroidal 2-point functions in free gauge theories); thus, the
non-vanishing Feynman diagrams in our theory map to a four
real-dimensional subspace of the decorated moduli space. Moreover,
since this subspace is clearly invariant under a rescaling of the
circumferences $p_i \to \alpha p_i$ (which corresponds to a
rescaling of all the Schwinger parameters, as mentioned in section
\ref{Preliminaries}), its projection onto the moduli space is at
most a three real-dimensional space (out of the four real
dimensions). So, again we find that the worldsheet correlation
function vanishes at generic points on the moduli space. In this
section we will investigate in detail the subspace on which the
correlation function is non-vanishing.
Some basic definitions and facts about elliptic functions which
will be used extensively in the following may be found in Appendix
\ref{Elliptic}.

\begin{figure}[htbp]
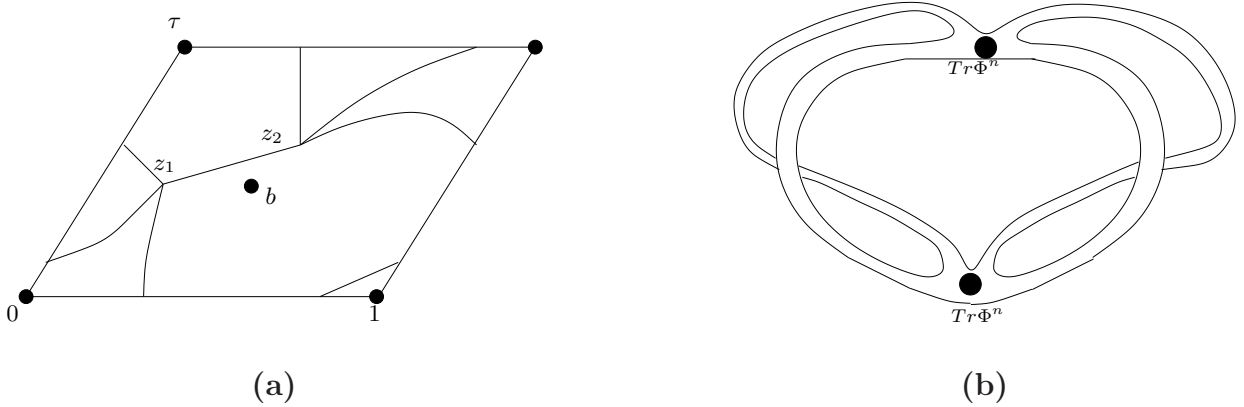

\begin{center}
$\begin{array}{c@{\hspace{1in}}c} \input{TorNSC.pstex_t} &
\input{TorNSC_FT.pstex_t}
 \\ [0.2cm]
\mbox{\bf (a)} & \mbox{\bf (b)}
\end{array}$
\caption{Two-point function on the torus without
self-contractions: {\bf (a)} The critical graph (dual to the gauge
theory graph). {\bf (b)} The gauge theory graph.} \label{NoSelf}
\end{center}
\end{figure}

The non-zero toroidal two-point functions in a free field theory
take the form
\begin{equation}
\langle \tr(\Phi^n(x_1))\tr(\Phi^n(x_2))\rangle_{T^2}.
\end{equation}
We are interested in diagrams without self-contractions. It is
easy to convince oneself that for such diagrams the two
circumferences, $p_0$ and $p_b$, must be equal, as depicted in
Figure \ref{NoSelf}(a). Any critical curve which is not topologically
of the form drawn in Figure \ref{NoSelf}(a) will have
self-contractions.

From the figure it is clear that the Strebel differential
corresponding to such a graph should have two double zeros (the
general two-point function on the torus will have four simple
zeros, but also self-contractions). The Strebel differential
should respect the periodicities of the torus and thus should be
an elliptic function.
For any elliptic function, the sum of the poles (weighted by their
order) minus the sum of the zeros (weighted by their order) is a
period of the torus. Denoting the zeros by $z_i$ ($i=1,2$), this
gives: \be 0+2b-\sum_{i=1}^4 ({\rm zeros})= 2b-2(z_1+z_2)=n\tau+m
\qquad (n,m\in \mathbb{Z}). \ee
There are now two distinct cases, one when the zeros and poles
satisfy
\be\label{t_cond1} b-z_1-z_2=
{\tilde n} \tau + {\tilde m},
\qquad ({\tilde n},{\tilde m} \in \mathbb{Z})\ee
and the other
where
\be \label{t_cond2} b-z_1-z_2\neq
{\tilde n} \tau + {\tilde m},
\qquad ({\tilde n},{\tilde m} \in \mathbb{Z}).\ee
\begin{figure}[htbp]
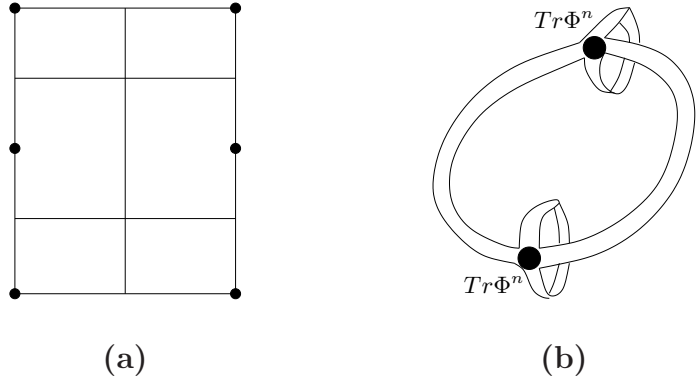

\begin{center}
$\begin{array}{c@{\hspace{1in}}c} \input{TorSC.pstex_t} &
\input{TorSC_FT.pstex_t}
 \\ [0.2cm]
\mbox{\bf (a)} & \mbox{\bf (b)}
\end{array}$
\caption{A specific two-point function on the torus with
self-contractions, with $\tau=2i$ and $b=i$: {\bf (a)} Critical
(dual) graph. {\bf (b)} Gauge theory graph.} \label{TorusSelfCon}
\end{center}
\end{figure}

Note that our demand that the differential will have two double
zeros is still not enough to rule out self-contractions. For
instance, the following differential (on a torus with $\tau=2i$)
has both double zeros and self-contractions, as depicted in Figure
\ref{TorusSelfCon}: \be q=-\frac{L^2}{4\pi^2}\wp(z\mid\tau=i)
dz^2. \ee Note that for this differential (which has $b=i$) we are
in the case (\ref{t_cond2}). In the rest of this section we will
analyze the general differential with no self-contractions, and we
will see that it satisfies (\ref{t_cond1}). Thus, we claim that
any diagram with self-contractions satisfies (\ref{t_cond2}), and
the ones without self-contractions satisfy (\ref{t_cond1}).

\begin{figure}[htbp]
\begin{center}
\input{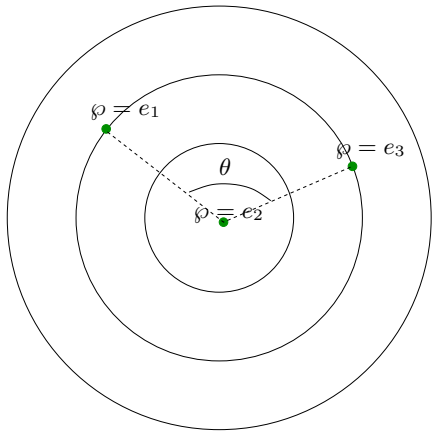}
\caption{The $\wp$-plane horizontal curves of (\ref{SpecDiff}).}
\label{CondTau}
\end{center}
\end{figure}

First, let us discuss a specific explicit example. We argue that
\be \label{SpecDiff}
q=-\frac{L^2}{4\pi^2}\{\wp(z)+\wp(z-\frac{\tau+1}{2})-e_1-e_3\}
dz^2 \ee is a good Strebel differential without self-contractions
for some class of $\tau$'s.
 This differential has double zeros at $z=\tau/2$ and $z=1/2$,
and double poles at $z=0$ and $z=(\tau+1)/2$.
  Thus, the zeros and poles satisfy (\ref{t_cond1}).
We can check if this is a Strebel differential by noting that \be
q=\left(\frac{iLd\wp}{4\pi(\wp-e_2)}\right)^2, \ee so we can
easily compute the edge lengths, say between the zeros $w_1=1/2$
and $w_3=\tau/2$: \be \sigma=\int_{w_1}^{w_3}\sqrt{q}=
\frac{iL}{4\pi}\int_{e_1}^{e_3}\frac{d\wp}{\wp-e_2}=\frac{iL}{4\pi}\ln
\frac{e_3-e_2}{e_1-e_2}. \ee For this to be real (note that this
is independent of the branch of the log) we should have
\be\label{t_condTau}\frac{e_3-e_2}{e_1-e_2}=e^{i\theta}\ee for
real $\theta$, which implies a condition on $\tau$. This condition
can be understood graphically : $e_3-e_2$ and $e_1-e_2$ should lie
on the same circle in the $\wp$ plane (see Figure \ref{CondTau}).
When this is satisfied, there will be two different edges of the
critical graph with lengths: \be
\sigma_1=L\frac{\theta(\tau)}{4\pi},\quad\quad \sigma_2=
L\frac{2\pi-\theta(\tau)}{4\pi}. \ee We note that the quantity
$\frac{e_3-e_2}{e_1-e_2}$ is used in the mathematical literature,
and is usually denoted by\footnote{ It satisfies
$k^2(\tau)=\theta^4_{10}(0|\tau)/\theta^4_{00}(0|\tau)$.} $k^2(\tau)$.
It satisfies that for any complex $a\neq0,1$ there is a $\tau$
such that $k^2(\tau)=a$. Thus, the solutions to (\ref{t_condTau})
give a one-dimensional subspace of the three-dimensional
submanifold we are looking for. One can show that the solution of
(\ref{t_condTau}) is simply (in the fundamental domain)
$|\tau|=1$ \footnote{This can be proven using elementary modular
properties of the so-called $\lambda$ functions.}. We will see
some examples of such solutions in numerical calculations below.
For these $\tau$'s,
the differential (\ref{SpecDiff}) is a perfectly good Strebel
differential, so we have found a class of examples of
differentials without self-contractions satisfying
(\ref{t_cond1}).

In the rest of this section we look for the most general Strebel
differential which has double zeros and satisfies (\ref{t_cond1}).
It
 can be written as:
 \begin{equation}\label{square reps.}
q=-
{c^2 \over 4\pi^2}\left(\frac{\wp(z-\frac{z_1+z_2}{2})-
\wp(\frac{z_1-z_2}{2})}{\wp(z-\frac{z_1+z_2}{2})-\wp(\frac{z_1+z_2}{2})}
\right)^2 dz^2,
\end{equation}
which has poles of second order at $0$, $b=z_1+z_2$ and zeros of
second order at $z_1$, $z_2$.
The equal residues are given by \be p_{0}^2=p_b^2 =c^2\left(
\frac{\wp(\frac{z_1+z_2}{2})-\wp(\frac{z_1-z_2}{2})}
{\frac{\partial\wp}{\partial z}(\frac{z_1+z_2}{2})} \right)^2. \ee
%
We fix the value of $\tau$,
and we wish to find $c$ and the zeros such
that (\ref{square reps.}) is a Strebel differential with a pole
at $b=z_1+z_2$. We have
solutions only for a three-dimensional submanifold of $\tau$, $b$,
determined by the reality of the integrals of $\sqrt{q}$
along curves $\gamma_i$
between the zeros :
\begin{align}\label{int}
    \s_i&={i c\over {2\pi}} \int_{\gamma_i} dz \frac{\wp(z-\frac{z_1+z_2}{2})-
\wp(\frac{z_1-z_2}{2})}{\wp(z-\frac{z_1+z_2}{2})-\wp(\frac{z_1+z_2}{2})}
= {i c\over {2\pi}} \int_{\gamma_i} dz\left(
1+\frac{\wp(\frac{z_1+z_2}{2})-\wp(\frac{z_1-z_2}{2})}
{\wp(z-\frac{z_1+z_2}{2})-\wp(\frac{z_1+z_2}{2})}\right)=\cr&=
{i c\over {2\pi}}\left(\gamma_i(1)-\gamma_i(0)+
\bigl(\wp(\frac{z_1+z_2}{2})-\wp(\frac{z_1-z_2}{2})\bigr)\int_{\gamma_i}
dz
\frac{1}{\wp(z-\frac{z_1+z_2}{2})-\wp(\frac{z_1+z_2}{2})}\right).
\end{align}
 This integral is expressible in terms of different
Weierstrass functions \be &&\int dz
\frac{1}{\wp(z-a)-\wp(a)}=\frac{\ln\frac{\s(z-2a)}{\s(z)}+2z\zeta(a)}{\wp'(a)},
\ee whose definitions are summarized in Appendix \ref{Elliptic}.
Denoting $\Delta z =z_2-z_1$, we find that the three independent
length-integrals lead to the equations :
\begin{itemize}
\item Direct $z_1\rightarrow z_2$ integral :
\begin{gather}\label{1-2}
\begin{align}
0&={\rm Re}\biggl\{ {c\over {2\pi}}\left(\Delta z+
\frac{\wp(\frac{z_1+z_2}{2})-\wp(\frac{\Delta
z}{2})}{\wp'(\frac{z_1+z_2}2)}\left(\ln\frac{\s(-z_1)}{\s(z_2)}-\ln\frac{\s(-z_2)}{\s(z_1)}+2\Delta
z\zeta(\frac{z_1+z_2}2)\right)\right) \biggr\}=\cr&=p_0 {\rm Re}\biggl\{
\frac{\wp'(\frac{z_1+z_2}2)\Delta
z}{\wp(\frac{z_1+z_2}{2})-\wp(\frac{\Delta z}{2})}+2\Delta
z\zeta(\frac{z_1+z_2}2) +2\ln\frac{\s(z_1)}{\s(z_2)} \biggr\},
\end{align}\end{gather}
\item $z_1\rightarrow z_2+1$ integral minus $z_1\rightarrow z_2$
integral :
\begin{gather}\label{1-2+1 - 1-2}
\begin{align}
0&=p_0{\rm Re}
\biggl\{\frac{\wp'(\frac{z_1+z_2}2)}{\wp(\frac{z_1+z_2}{2})-\wp(\frac{\Delta
z}{2})}+2\zeta(\frac{z_1+z_2}2)+
\ln\frac{\s(-z_1+1)}{\s(z_2+1)}-\ln\frac{\s(-z_1)}{\s(z_2)}\biggr\}=\cr&=
p_0
{\rm Re}
\biggl\{\frac{\wp'(\frac{z_1+z_2}2)}{\wp(\frac{z_1+z_2}{2})-\wp(\frac{\Delta
z}{2})}+2\zeta(\frac{z_1+z_2}2)-2(z_1+z_2)\eta_1 \biggr\},
\end{align}\end{gather}
\item $z_2\rightarrow z_1+\tau$ integral minus $z_2\rightarrow z_1$
integral :
\begin{gather}\label{2-1+t - 2-1}
\begin{align}
0&= p_0
{\rm Re}\biggl\{
\frac{\wp'(\frac{z_1+z_2}2)\tau}{\wp(\frac{z_1+z_2}{2})-\wp(\frac{\Delta
z}{2})}+2\tau\zeta(\frac{z_1+z_2}2)-2(z_1+z_2)\eta_3
\biggr\}=\cr&=p_0
{\rm Re}\biggl\{\frac{\wp'(\frac{z_1+z_2}2)\tau}{\wp(\frac{z_1+z_2}{2})-
\wp(\frac{\Delta z
}{2})}+2\tau\zeta(\frac{z_1+z_2}2)-2(z_1+z_2)\tau\eta_1
+2i(z_1+z_2)\pi \biggr\}.
\end{align}
\end{gather}
\end{itemize}
It is possible to combine the last two equations into
\begin{equation}
    \frac{\wp'(\frac{z_1+z_2}2)}{\wp(\frac{z_1+z_2}{2})-\wp(\frac{\Delta
z}{2})}+2\zeta(\frac{z_1+z_2}2)-2(z_1+z_2)\eta_1=-2i\pi {\rm
Im}(z_1+z_2)/\tau_2,
\end{equation}
which is further conveniently rewritten as
\begin{equation}
\label{convenient}
    \wp(\frac{\Delta
z}{2})=\frac{\wp'(\frac{z_1+z_2}2)}{2\zeta(\frac{z_1+z_2}2)-2(z_1+z_2)\eta_1
+2i\pi {\rm Im}(z_1+z_2)/\tau_2}+\wp(\frac{z_1+z_2}{2}).
\end{equation}
An immediate consequence is that for small $b=z_1+z_2$ we have
\begin{equation}
    \wp(\frac{\Delta
z}{2})\sim \biggl(-2\eta_1+\frac{2\pi
i}{\tau_2}\frac{{\rm Im}(b)}{b}\biggr) + O(b)=
 O(1),
\end{equation}
which means that both of the zeros approach a constant (depending
on the phase of $b$) as $b\to 0$.

\begin{figure}[htbp]
\begin{center}
$\begin{array}{ccc} \epsfig{file=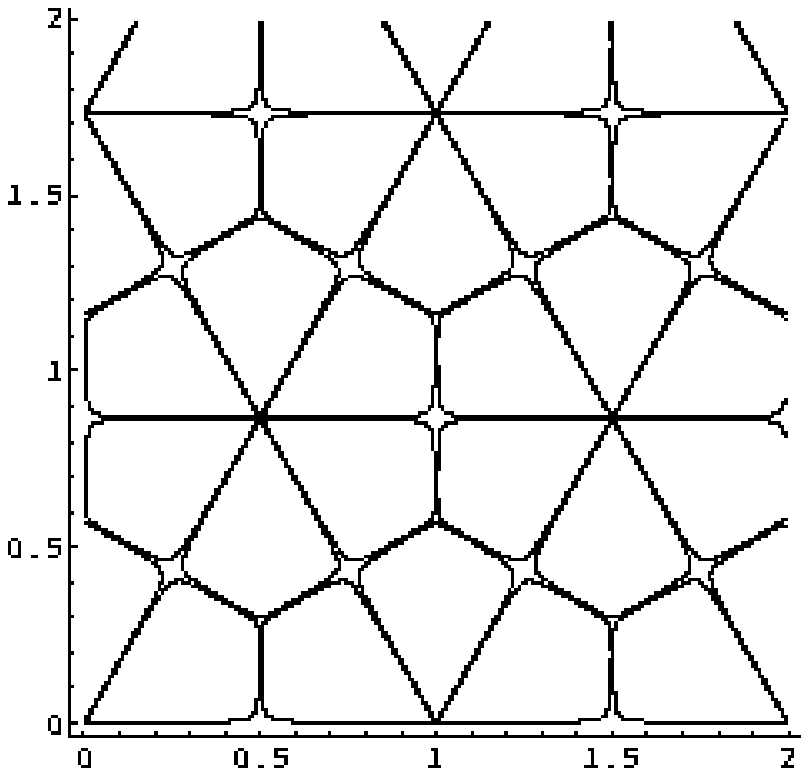, width=1.7in} &
\epsfig{file=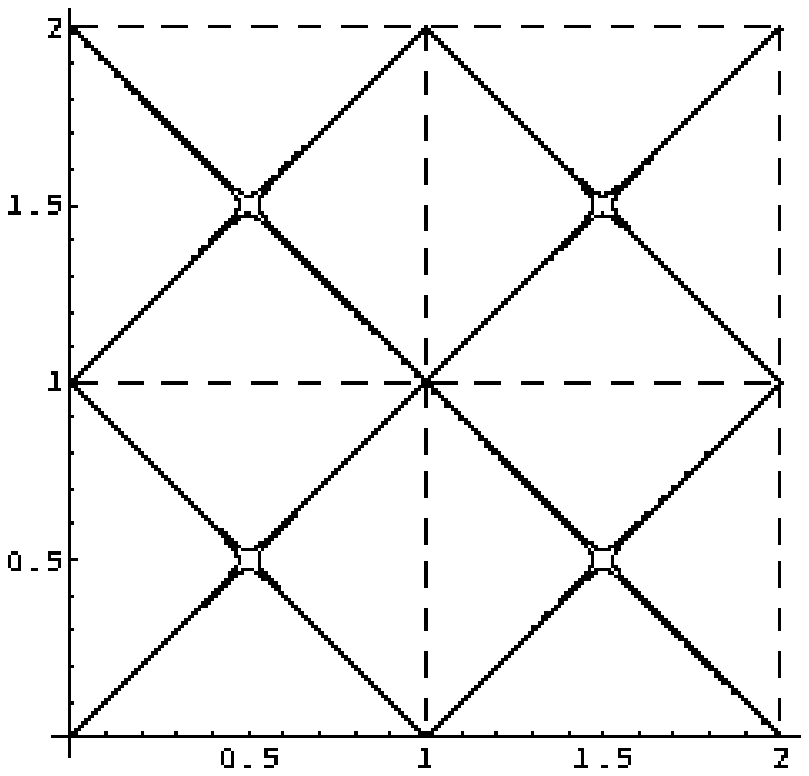, width=1.7in} &
\epsfig{file=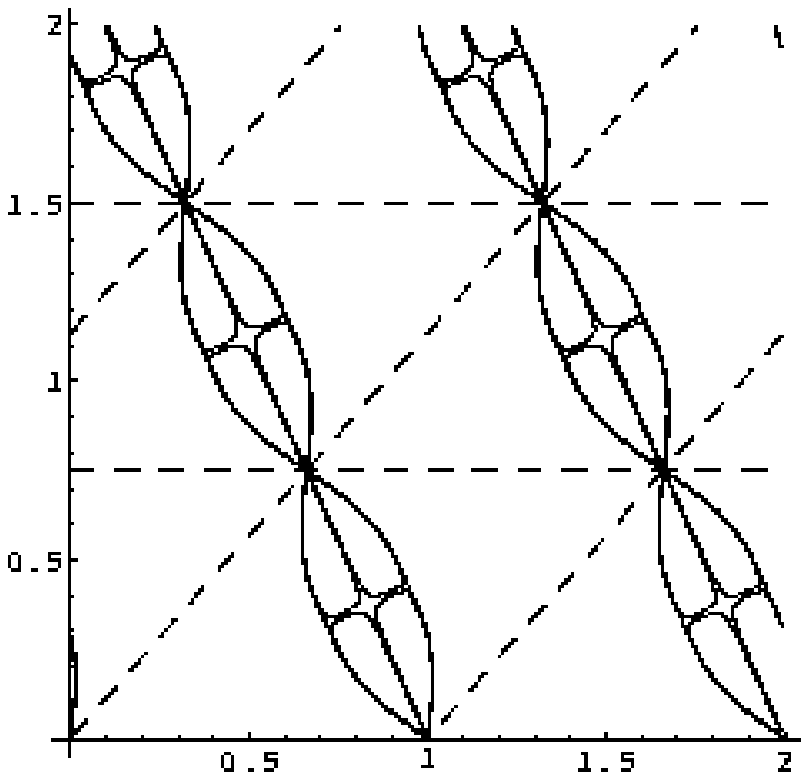, width=1.7in}
 \\ [0.2cm]
\mbox{\bf (a)} & \mbox{\bf (b)} & \mbox{\bf (c)}\\ [0.2cm]
\epsfig{file=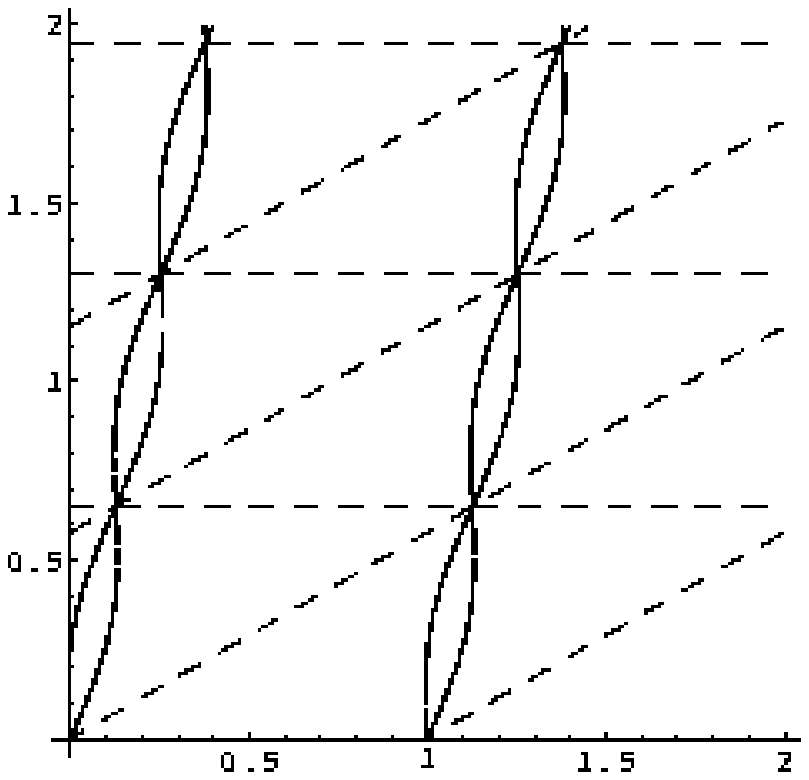, width=1.7in} &
\epsfig{file=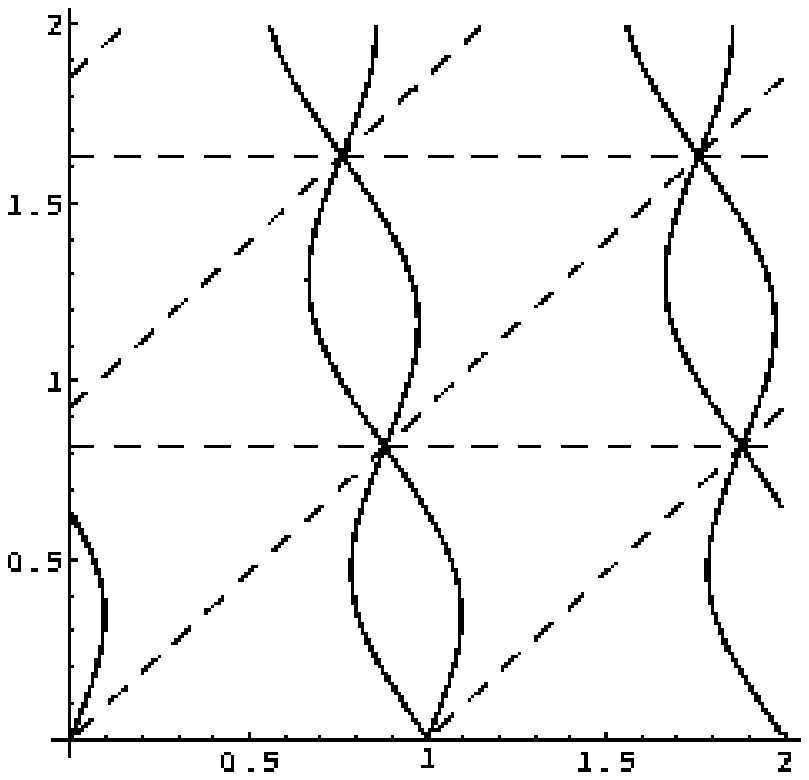, width=1.7in} &
\epsfig{file=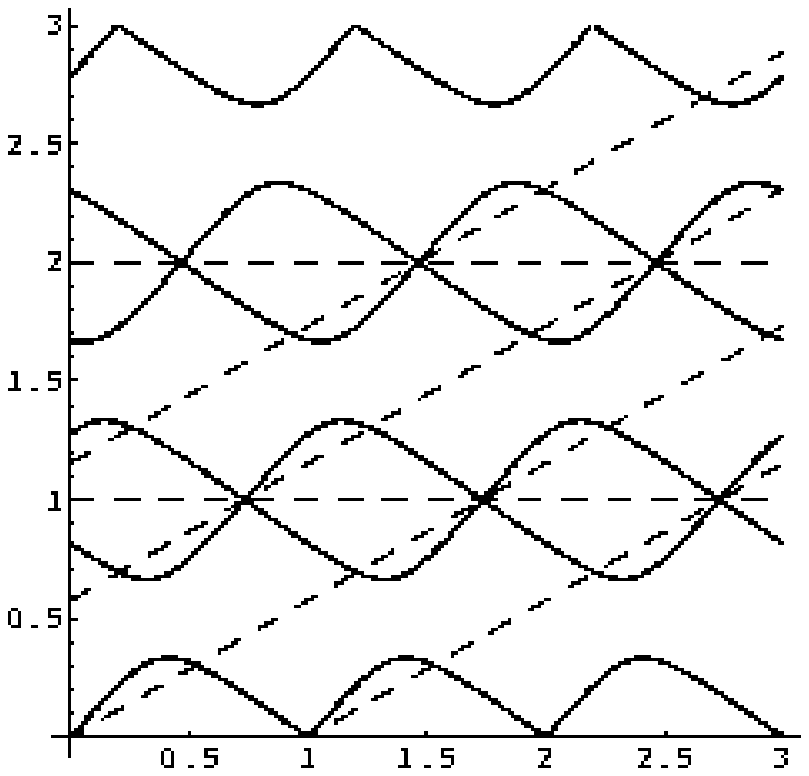, width=1.7in}
 \\ [0.2cm]
\mbox{\bf (d)} & \mbox{\bf (e)} & \mbox{\bf (f)} \\ [0.2cm]
\end{array}$
\caption{Numerical plots of
the curves on which the second insertion, $b$, can lie for
six different values of $\tau$: {\bf (a)}~$\tau=e^{\pi i/3}$, {\bf
(b)}~$\tau=i$, {\bf (c)}~$\tau=e^{i\pi/3.7}$,
{\bf (d)}~$\tau=1.3e^{i\pi/6}$, {\bf (e)}~$\tau=1.2e^{i\pi/4.2}$,
{\bf (f)}~$\tau=2e^{i\pi/6}$. The fundamental domain is plotted with dashed
lines (in figure (a) $b$ can lie on the boundaries of the
fundamental domain we chose). The small ``diamonds'' appearing in
(a),(b),(c) are the result of numerical errors, and should be
replaced by straight lines going through the middle of the
``diamonds''.} \label{TorNum}
\end{center}
\end{figure}

We define a function $K(z_{1,2},{\bar z_{1,2}},\tau,{\bar \tau})$
by the inverse Weierstrass function acting on (\ref{convenient}) :
\begin{equation}\label{zerodiff}
    \frac{\Delta
z}{2}=\wp^{-1}\left(\frac{\wp'(\frac{z_1+z_2}2)}
{2\zeta(\frac{z_1+z_2}2)-2(z_1+z_2)\eta_1
+2i\pi {\rm
Im}(z_1+z_2)/\tau_2}+\wp(\frac{z_1+z_2}{2})\right)\equiv
K(z_{1,2},\bar{z}_{1,2},\tau,\bar{\tau}).
\end{equation}
Plugging this result back into (\ref{1-2}) we get an equation
directly for $z_{1,2}$ so that the differential (\ref{square
reps.}) is a Strebel differential on the torus with modulus $\tau$
and with a marked point at
$(z_1+z_2)\ \text{mod}(\text{lattice})$ :
\begin{gather}\label{1-2new}
0={\rm Re}\Biggl\{ \biggl(2(z_1+z_2)\eta_1 -{2i\pi\over \tau_2} {\rm
Im}(z_1+z_2)\biggr)K(z_{1,2},\bar{z}_{1,2},\tau,\bar{\tau})
+\ln\frac{\s(\frac{z_1+z_2}{2}-K(z_{1,2},\bar{z}_{1,2},\tau,\bar{\tau}))}
{\s(\frac{z_1+z_2}{2}+K(z_{1,2},\bar{z}_{1,2},\tau,\bar{\tau}))}
\Biggr\}.
\end{gather}
Note that, as expected, the definition of $K$ and equation
(\ref{1-2new}) depend only on $z_1+z_2$, so this gives a real
equation for $b$ and $\tau$; however, we chose to write the
equation as an equation for the $z_i$ and not for $b$ because it
is periodic under simultaneous shifts of $z_1$ and $z_2$ by
$n+m\tau$, but not under general lattice shifts of $z_1+z_2$
(although the space of solutions for $b=z_1+z_2$ is of course
periodic).
In practice, we solve for the sum of the zeros and
then $b$ is uniquely determined, modulo the
lattice.
We are now able
to find directly
the curves on which $b$ can lie.
We expect that for a generic $\tau$ the insertion $b$ can be moved
on some curves in the fundamental domain. In Figure \ref{TorNum}
we show these curves (computed numerically from (\ref{1-2new}))
for some representative values of $\tau$.
For some special values of $\tau$ which have symmetry we have
non-generic behaviour. One can see that the point $b=(1+\tau)/2$
is indeed a solution when $|\tau|=1$, as discussed above.

The equations above simplify when we have only three edges in the
field theory graph (which is to be drawn on the torus). Here, the
dual graph has only one zero of order four, which we denote by
$z_0$, and the pole is at $b=2z_0$. One can read off the
constraint on the moduli space of marked tori from
(\ref{zerodiff}) with $\Delta z = 0$:
\be \label{subspacecons} \zeta(z_0)-2z_0\eta_1 +2i\pi {\rm
Im}(z_0)/\tau_2 =0.\ee

Note that we have to look for solutions in which $b=2z_0\
\text{mod (lattice)}$ is not a lattice point (since these do not
correspond to Strebel differentials). This defines some two
dimensional subspace of the moduli space, which we can analyze
numerically. We observe that there are no solutions for $\tau=i$,
for instance. In addition, for $\tau=e^{\pi i /3}$ one can see
that $z_0=(1+e^{\pi i /3})/3$ and $z_0=2(1+e^{\pi i /3})/3$ are
solutions of equation (\ref{subspacecons}), with $b=2(1+e^{\pi i
/3})/3$ and $b=(1+e^{\pi i /3})/3$, respectively\footnote{ Note
that other values of $z_0$ for which $2z_0$ equals (modulo the
lattice) these values of $b$ are not solutions, so there is indeed
a
unique Strebel differential
on a given marked Riemann surface.}. These points appear in
Figure \ref{TorNum}(a) as the junction points where three lines meet.
This is a general phenomenon: for all values of $\tau\neq i$ with
$|\tau|=1$ there are junctions of three lines which
correspond to one edge going to zero length.

\vskip 3.0cm \centerline{\bf Acknowledgements} We would like to
thank  O. Bergman, M. Berkooz, N. Itzhaki, Y. Oz, D. Reichmann, A.
Schwimmer, and especially R. Gopakumar for useful discussions. The
work of OA and ZK was supported in part by the Israel-U.S.
Binational Science Foundation, by the Israel Science Foundation
(grant number 1399/04), by the Braun-Roger-Siegl foundation, by
the European network HPRN-CT-2000-00122, by a grant from the
G.I.F., the German-Israeli Foundation for Scientific Research and
Development, by Minerva, and by the Einstein center for
theoretical physics. The work of SSR is supported in part by the
Israel Science Foundation under grant no. 568/05.
\appendix
\section*{Appendix}
\section{More Sphere Diagrams}\label{SphereApp}
\begin{figure}[htbp]
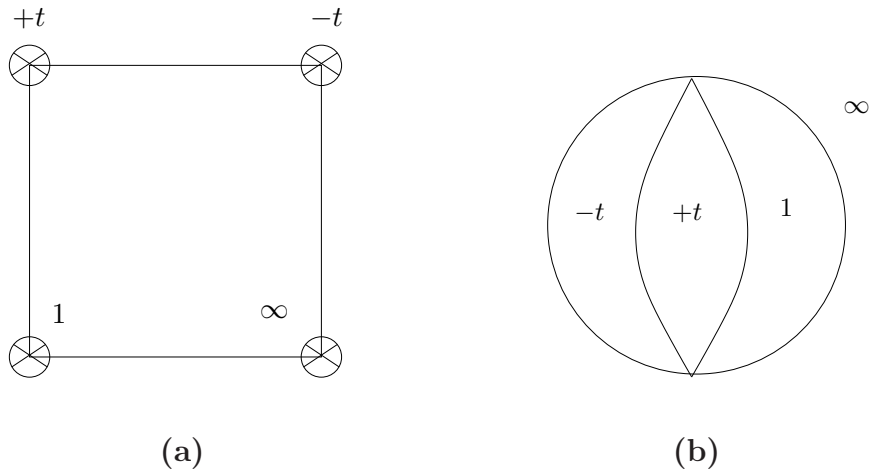

\begin{center}
$\begin{array}{c@{\hspace{1in}}c}
\input{4pointCirc.pstex_t} &\input{4pointCircD.pstex_t}
 \\ [0.4cm]
\mbox{\bf (a)} & \mbox{\bf (b)}
\end{array}$
\caption{Four-point circle diagram on a sphere : {\bf (a)} Gauge theory
graph, {\bf (b)} Dual graph.} \label{4pointCircle}
\end{center}
\end{figure}

In this appendix we will discuss two additional sphere diagrams,
the four-point and five-point diagrams with the topology of a
circle (with edges running all around the circle). Such circular
diagrams are the only contributions to correlation functions of
the form $\langle\prod_{i=1}^n \tr(\Phi^2(x_i))\rangle$. Our
discussion will illustrate further some of the points which were
mentioned in section \ref{Sphere}. The four-point gauge theory
diagram and dual critical graph are drawn in Figure
\ref{4pointCircle}. In the four-point diagram the decorated moduli
space contains one modulus $t$ and four circumferences; however,
from the diagram it is clear that there is one linear relation
between the circumferences (in the specific case drawn in Figure
\ref{4pointCircle}(b) it is $p_t+p_{\infty}= p_1+p_{-t}$), so the
diagram maps at most to a five-dimensional subspace of the
decorated moduli space. However, in the gauge theory we have only
four Schwinger parameters, so the diagram will localize on a
codimension-one subspace of this five-dimensional subspace,
similar to what we found in previous cases.
In the case of the $X$ amplitude which we discussed in section
\ref{SphereX} the $\delta$-function describing the localization
did not depend on the magnitude of the complex modulus $t$ (at
leading order for small $t$), so the question of the leading power
appearing in the OPE was well-defined. On the other hand, for the
circle diagram we will see that the constraint does depend on
$|t|$ even in the leading order, and thus we cannot determine the
powers appearing in the OPE unless we explicitly perform the
integrals over the circumferences (which we were unable to do). In
the case of the five-point circle diagram, we will see that the
string theory diagram can be written without any
$\delta$-functions. We can translate explicitly all the
circumferences to the string moduli (and one overall scaling
parameter).
\subsection{Circular four-point function}
\begin{figure}[h]
\begin{center}
$\begin{array}{c@{\hspace{1in}}c}
\epsfxsize=1.6in \epsffile{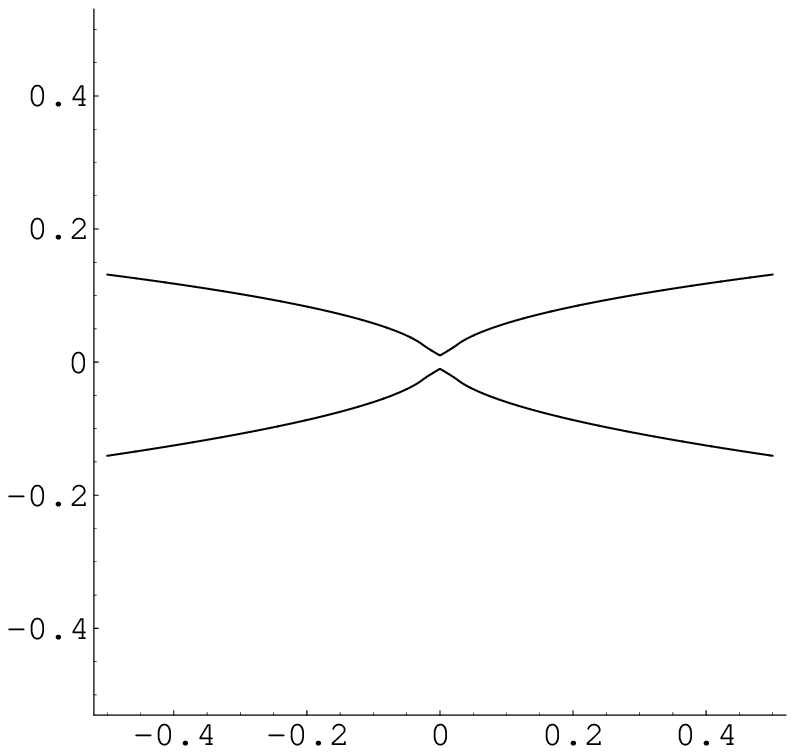} &
    \epsfxsize=1.6in
    \epsffile{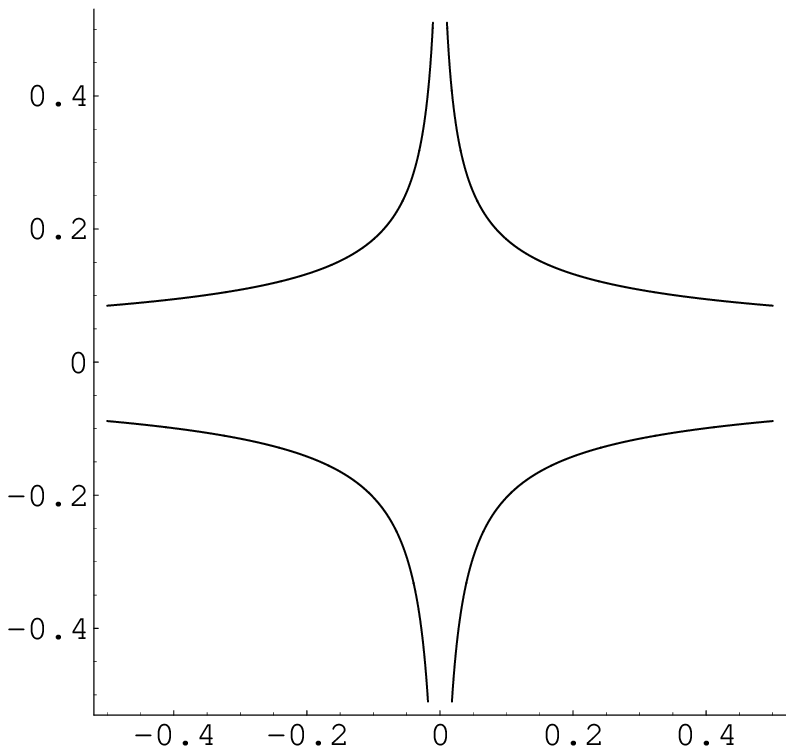} \\ [0.4cm]
\mbox{\bf (a)} & \mbox{\bf (b)}
\end{array}$
\end{center}
\caption{Solutions to the constraint (\ref{fourCircleconstraint}), for a
specific value $t=0.0005e^{2.75i\pi}$ :
{\bf (a)} This graph shows $B(A,t)$,
{\bf (b)} This graph shows $B(1/A,t)$.
We see that the constraint can be solved to obtain as the
three free real parameters the complex $t$ and either $A$ or $B$.}
\label{figtest-fig}
\end{figure}

The most general Strebel differential
for the circular four-point diagram (with insertions at
$z=1,\pm t,\infty$) has the following form:
\be
\label{fourcircleStrebel}
q=-\frac{1}{4\pi^2}\left(\gamma_+\frac{p_{t}}{z-t}+
\gamma_-\frac{p_{-t}}{z+t}+\gamma_1\frac{p_{1}}{z-1}\right)^2dz^2,
\ee
where the $\gamma_i$ are sign choices which are determined by the
ordering of the insertions around the circle. The ordering specifies
a precise relation between the circumferences, which we wrote as
$\gamma_+p_t+\gamma_-p_{-t}+\gamma_1p_1=p_\infty$. In order to check
that (\ref{fourcircleStrebel}) is a Strebel differential
we need to make sure that the edge-lengths are real.
For given circumferences we will find one
such condition, which implies that (for given circumferences)
$t$ cannot take any
complex value but lies in a one-dimensional subspace.
The condition is :
\be
\label{fourCircleconstraint}
0&=&{\rm Im}\{\int_{c_+}^{c_-}dz\sqrt{q}\}={\rm Re}\{\int_{c_+}^{c_-}
\frac{dz}{2\pi}
\biggl(\gamma_+\frac{p_{t}}{z-t}+\gamma_-\frac{p_{-t}}{z+t}+
\gamma_1\frac{p_{1}}{z-1}
\biggr)\}=\\
&=&\frac{1}{2\pi}{\rm Re}\biggl(\gamma_+p_{t}\ln(z-t)+\gamma_-p_{-t}\ln(z+t)+
\gamma_1p_1\ln(z-1)\biggr)\bigg|_{c_+}^{c_-},\nonumber
\ee
where $c_{\pm}$ are the two zeros of the differential; defining
\be
B\equiv \frac{\gamma_+p_t+\gamma_-p_{-t}}{p_\infty},\quad\quad
A\equiv \frac{\gamma_-p_{-t}-\gamma_+p_t}{p_\infty},
\ee
they are given by:
\be
c_\pm=\half\biggl(B+At\pm\sqrt{(B+At)^2-4t(A+(B-1)t)}\biggr).
\ee
Equation (\ref{fourCircleconstraint})
is an explicit real constraint on the complex parameter $t$ and
on the real parameters $A$ and $B$. A numerical solution of this
constraint is depicted in Figure \ref{figtest-fig} for a specific
value of $t$.
The solutions $B(A,t)$ (or $A(B,t)$) scale with $t$ even at the
leading order of small $t$. Thus, before solving the constraint and
integrating over the circumferences (which we were not able to do) we cannot
make any claim about the OPE arising from this diagram.
\subsection{Circular five-point function}
We turn now to the circular five-point diagram.
The field theory and the dual graphs are very
similar to the four-point case we discussed above. However,
a crucial difference is that in this case there is no constraint
relating the circumferences; in fact, there is a simple invertible
linear relation between the five edge-lengths and the five circumferences.
Choosing the insertions to lie at $z=1,\pm t,b,\infty$,
the general Strebel differential has the form:
\be
q=\left( \frac{ip_\infty}{2\pi}\frac{(c_1-z)^{3/2}(c_2-z)^{3/2}}
{(1-z)(t^2-z^2)(b-z)}dz\right)^2.
\ee
On the gauge theory side we have five Schwinger parameters; on the string
theory side we have
five independent circumferences and two complex moduli ($t$ and $b$).
Recalling that the moduli do not change when we rescale all the edges,
we expect to be able to
rewrite the gauge theory diagram as an integral over all the moduli and over
an overall scale factor related to scaling all the circumferences.
Namely, we expect that for any set of
circumferences (up to rescalings) which is consistent with the
topology of the graph, we will find a single value for the moduli.

The technical details are as follows. First, we change variables
from the circumferences $p_i$ to an overall scaling parameter
$p_{\infty}$ and four parameters $\alpha_i$ related to ratios of
the circumferences, defined by :
  \be
\label{alphadef}
\frac{p_1}{p_\infty}&=&(-1)^{\gamma_1}\frac{(c_1-1)^{3/2}(c_2-1)^{3/2}}
{(1-t^2)(b-1)}\equiv \a_1^{3/2},\\
\frac{p_{\pm }}{p_\infty}&=&(-1)^{\gamma_\pm}\frac{(c_1\mp t)^{3/2}
(c_2\mp t)^{3/2}}{2t(1\mp t)(b\mp t)}\equiv\a_\pm^{3/2},\nonumber\\
\frac{p_{b }}{p_\infty}&=&(-1)^{\gamma_b}\frac{(c_1-
b)^{3/2}(c_2- b)^{3/2}}{(t^2- b^2)(1-
b)}\equiv\a_b^{3/2}.\nonumber \ee
Here, we denoted the residues at $\pm t$ by $p_{\pm}$. The sign parameters
 $\gamma_i$ take values in $\{0,1\}$, according to which region of the
parameter space we are in
(as for the four-point function).
The four complex
 equations (\ref{alphadef}) give a relation between the (real parameters)
$\alpha_i$, the two complex zero positions $c_i$
 and the two complex moduli. We would like to write the four
 circumference parameters $\a_i$ as functions of the moduli
 space. Define four quantities which depend only on the moduli:
 \be
F_\pm &\equiv&e^{\frac{4\pi i}{3}\eta_\pm+\frac{2\pi i}{3}(1-\gamma_{\pm})}\biggl(2t(1\mp t)(b\mp t)\biggr)^{2/3},\\
F_b &\equiv&e^{\frac{4\pi i}{3}\eta_b+\frac{2\pi
i}{3}(1-\gamma_{b})}\biggl((t^2- b^2)(1- b)\biggr)^{2/3},\nonumber \\
F_1 &\equiv& e^{\frac{4\pi i}{3}\eta_1+\frac{2\pi
i}{3}(1-\gamma_{1})}\biggl((t^2- 1)(1- b)\biggr)^{2/3},\nonumber
 \ee
where the $\eta_i$ can take the values $\{0,1,2\}$
(assuming a specific choice of phase for the $2/3$ power).
From the definition of
$\a_\pm$ we find :
 \be
 c_1c_2&=&\half\biggl(\a_+F_++\a_-F_--2t^2\biggr),\\
 c_1+c_2&=&\frac{1}{2t}\biggl(-\a_+F_++\a_-F_-\biggr).\nonumber
 \ee
%
%
Now, we have
\be
\a_b&=&\frac{b^2-t^2}{F_b}+\a_+\biggl(1+\frac{b}{t}\biggr)\frac{F_+}{2F_b}+\a_-\biggl(1-\frac{b}{t}\biggr)\frac{F_-}{2F_b},\\
\a_1&=&\frac{1-t^2}{F_1}+\a_+\biggl(1+\frac{1}{t}\biggr)\frac{F_+}{2F_1}+\a_-\biggl(1-\frac{1}{t}\biggr)\frac{F_-}{2F_1},\nonumber
\ee
and we have two more equations which come from the reality of the
circumferences:
\be
-{\rm Im}\left(\frac{b^2-t^2}{F_b}\right)&=&\a_+{\rm Im}\left(\bigl(1+
\frac{b}{t}\bigr)\frac{F_+}{2F_b}\right)
+\a_-{\rm Im}\left(\bigl(1-\frac{b}{t}\bigr)\frac{F_-}{2F_b}\right),\\
-{\rm Im}\left(\frac{1-t^2}{F_1}\right)&=&\a_+{\rm Im}\left(\bigl(1+
\frac{1}{t}\bigr)\frac{F_+}{2F_1}\right)
+\a_-{\rm Im}\left(\bigl(1-\frac{1}{t}\bigr)\frac{F_-}{2F_1}\right).\nonumber
\ee
We can now write $\a_{\pm}$ purely as a function of the circumferences :
 \be
 \half\a_+&=&\frac{-{\rm Im}\biggl[\frac{b^2-t^2}{F_b}\biggr]
{\rm Im}\biggl[\biggl(1-\frac{1}{t}\biggr)\frac{F_-}{F_1}\biggr]+
 {\rm Im}\biggl[\frac{1-t^2}{F_1}\biggr]{\rm Im}\biggl[\biggl(1-
\frac{b}{t}\biggr)\frac{F_-}{F_b}\biggr]}
{{\rm Im}\biggl[\biggl(1+\frac{b}{t}\biggr)\frac{F_+}{F_b}\biggr]
{\rm Im}\biggl[\biggl(1-\frac{1}{t}\biggr)\frac{F_-}{F_1}\biggr]
-{\rm Im}\biggl[\biggl(1+\frac{1}{t}\biggr)\frac{F_+}{F_1}\biggr]
{\rm Im}\biggl[\biggl(1-\frac{b}{t}\biggr)\frac{F_-}{F_b}\biggr]},\\
\half\a_-&=&\frac{{\rm Im}\biggl[\frac{b^2-t^2}{F_b}\biggr]
{\rm Im}\biggl[\biggl(1+
\frac{1}{t}\biggr)\frac{F_+}{F_1}\biggr]-
 {\rm Im}\biggl[\frac{1-t^2}{F_1}\biggr]{\rm Im}\biggl[\biggl(1+
\frac{b}{t}\biggr)\frac{F_+}{F_b}\biggr]}
{{\rm Im}\biggl[\biggl(1+\frac{b}{t}\biggr)
\frac{F_+}{F_b}\biggr]{\rm Im}\biggl[\biggl(1-\frac{1}{t}\biggr)
\frac{F_-}{F_1}\biggr]
-{\rm Im}\biggl[\biggl(1+\frac{1}{t}\biggr)\frac{F_+}{F_1}\biggr]
{\rm Im}\biggl[\biggl(1-\frac{b}{t}\biggr)\frac{F_-}{F_b}\biggr]}.\nonumber
 \ee
Thus, we have found explicitly the dictionary from the moduli to the ratios of
the circumferences.
We can calculate the
Jacobian in a straightforward way, though the explicit expression is
quite complicated.
The only things which are not set yet are the phases
$\gamma_i$ and $\eta_i$; this is a discrete choice of parameters,
depending on which
region of the moduli space we are in. The exact regions are not
simple to compute, but we will mostly be interested in the OPE
region $t\to 0$ where we will be able to determine the phases.

The gauge theory result for a five-point function
$\vev{\prod_{i=1}^5 \tr(\Phi^{J_i}(x_i))}$ is given by:
 \be
 G\propto \int (\prod_{i=1}^5d\s_i\s_i^{m_i-1})
e^{-\frac{(x_1-x_2)^2\s_1}{4}-\frac{(x_2-x_3)^2\s_2}{4}
-\frac{(x_3-x_4)^2\s_3}{4}-\frac{(x_4-x_5)^2\s_4}{4}
-\frac{(x_5-x_1)^2\s_5}{4}},
 \ee
where the $m_i$ are related in a simple way to the $J_i$.
There is a similar
simple relation between the $\s_i$ and the circumferences $p_i$
(written in a specific ordering depending on the ordering of the vertices
around the graph). Defining $\b_i\equiv p_i/p_\infty$, it is given by
\be
\label{sigmabetarel}
\s_i=\half \sum_{k=0}^4 (-1)^kp_{i+k ({\rm mod\ } 5)}=
\half p_\infty\sum_{k=0}^4 (-1)^k\b_{i+k ({\rm mod\ } 5)}.
\ee
We can now change coordinates from the $\s_i$ to four $\b_i$ and $p_\infty$.
The Jacobian is proportional to $p_\infty^4$. Defining
\be
G_2(\b)=
\frac{(x_1-x_2)^2\s_1+(x_2-x_3)^2\s_2+(x_3-x_4)^2\s_3+(x_4-x_5)^2\s_4+(x_5-x_1)^2\s_5}{4p_{\infty}}\ee
(which is implicitly a function of the $\beta$'s using
(\ref{sigmabetarel})), the amplitude becomes:
 \be
\label{fivepointbeta}
 G&\propto&\int_0^\infty dp_\infty p_\infty^{4+\sum_{i=1}^5(m_i-1)}
\int_0^\infty\prod_{i=1}^4 d\b_i \biggl[\prod_{j=1}^5\biggl(\half\sum_{k=0}^4
(-1)^k\b_{j+k}\biggr)^{m_j-1}\biggr]e^{-p_\infty G_2(\b)}\\
  &\propto&\int_0^\infty\prod_{i=1}^4 d\b_i \frac{G_1(\b)}
{(G_2(\b))^{\sum_{i=1}^5m_i}},\nonumber
\ee
where
\be
G_1(\b)\equiv \prod_{j=1}^5\biggl(\half\sum_{k=0}^4
(-1)^k\b_{j+k}\biggr)^{m_j-1}.
 \ee
Above we calculated the dictionary between the $\a_i=\b_i^{2/3}$ and
the moduli $t,b$, so we can now rewrite (\ref{fivepointbeta})
as a (complicated) integral over the moduli.

All that remains to compute the OPE is to analyze the choice of
phases $\eta_i$, $\gamma_i$. We assume that the points $\pm t$ lie
in adjacent disks of the critical graph; specifically, we choose
the cyclic order of the points to be $t,-t,1,b,\infty$.
For a ``generic'' choice of signs, with different signs at
$t$ and $-t$, we find that as $t\to 0$
$\a_\pm\sim |t|^{1/3}$ and $\a_{b,1}\sim |t|^0$, which implies that
$\b_\pm\sim\sqrt{|t|}$ and $\b_{b,1}$ are finite.  Since each
circumference $p_i$ is the sum of two adjacent $\sigma$'s, this means
that three $\s_i$'s go as $\sqrt{|t|}$ and two go as $|t|^0$.
We claim that this case is inconsistent.
If we have
three edges going to zero
then we should have that (at leading order in $t$)
$p_b=p_1+p_\infty$, or $\alpha_b^{3/2}=\alpha_1^{3/2}+1$.
In the limit of $t\to 0$ we have:
 \be
 F_\pm=\delta_\pm\biggl(2tb\biggr)^{2/3},\quad F_b=\delta_b\biggl(b^2(1-b)\biggr)^{2/3},\quad F_1=\delta_1\biggl(1-b\biggr)^{2/3},
 \ee
and
%
%
%
 \be
2\a_b&=&\frac{2b^2}{F_b}+\biggl(F_+\a_+-F_-\a_-\biggr)\frac{b}{tF_b},\\
2\a_1&=&\frac{2}{F_1}+\biggl(F_+\a_+-F_-\a_-\biggr)\frac{1}{tF_1}.\nonumber
\ee The only solution we could find to\footnote{
We can prove this is the only solution in some regions of parameter space,
such as $1/|t|\gg|b|\gg1$.}
$\a_b^{3/2}=\a_1^{3/2}+1$ is to choose $\delta_+=\delta_-$; in
this case at leading order we have $F_+=F_-$ and $\a_+=\a_-$ so
$F_+\a_+-F_-\a_-=0$, and then
\be
\a_b^{3/2} - \a_1^{3/2}=
\biggl(\frac{b^3}{\delta_b^{3/2}b^2(1-b)} -
\frac{1}{\delta_1^{3/2}(1-b)}\biggr),
\ee
which indeed equals one for an appropriate choice of $\delta_b$,$\delta_1$.

However, in this case of $\delta_+=\delta_-$ it actually turns out
that the scaling at small $t$ is different. In this case we find
$\a_\pm\sim |t|^{-2/3}$ and $\a_{b,1}\sim |t|^0$, which implies that
$\b_\pm\sim 1/|t|$ and $\b_{b,1}\sim 1$. Thus, one of the $\s_i$'s (the
edge connecting the two operators which are coming together) scales
as $1/|t|$ while the others are finite (as we found in some of the
previous OPEs we analyzed).
The Jacobian for the change of variables from the $\beta$'s to the
absolute value and phase of $t$ and $b$ goes in this case
as $A/|t|^{2}+B/|t|$ for some constants $A$ and $B$.
It is easy to see that $G_1 \propto |t|^{1-\frac{1}{2}(J_t+J_{-t}-J_1+J_b
-J_{\infty})}$ while $G_2 \propto (x_t-x_{-t})^2/|t|$, so we find that
the full diagram scales as (for small $t$)
\begin{align}
\label{fiveOPE}
G & \propto \int d|t| d(\arg(t))
\frac{1}{|t|^2} |t|^{1-\frac{1}{2}(J_t+J_{-t}-J_1+J_b
-J_{\infty})} (|t|/(x_t-x_{-t})^2)^{\sum_i m_i} \cr & \propto
\int d^2t |t|^{J_1+J_{\infty}-2} |x_t-x_{-t}|^{-\sum_i J_i}.
\end{align}
It is easy to see that whenever the diagram is non-zero, the power
of $t$ that we find is always larger than the power we found in
the $star$ diagrams which was $|J_t-J_{-t}|-2$. So, the OPE is
consistent with this being the minimal power appearing in the OPE,
and we see that in many cases in the five-point function all the
leading terms in the OPE cancel and the first term appearing is
(\ref{fiveOPE}). The space-time dependence of the leading small
$t$ result is again consistent with scaling invariance.

\section{A Short Primer on Elliptic Functions}\label{Elliptic}

In our study of Strebel differentials on a torus we need to use
elliptic functions. In this section we will briefly review the
basic theory of such functions and some useful facts. A
meromorphic function $f(z)$ which is doubly periodic, \be
f(z+a)=f(z+\tilde a)=f(z), \ee where the periods satisfy ${\rm
Im}({\tilde a}/a)>0$, is called an elliptic function. These
functions have many nice properties, including :
\begin{itemize}
\item The sum of the residues of the simple poles of an elliptic function
inside the period-parallelogram (the parallelogram spanned by the two
periods $a$ and $\tilde a$) is equal to zero.
\item The number of zeros of a non-constant elliptic function inside the
period parallelogram is equal to the number of poles, where the
zeros and poles are weighted by their degrees.
\item The sum of the positions of the zeros of a non-constant elliptic
function inside the
period parallelogram differs from the sum of the position of the
poles by a period (an integer linear combination of $a$ and $\tilde a$),
where again the sum is weighted by the degrees of
the zeros and poles.
\end{itemize}

A useful example of an elliptic function is the Weierstrass
function, which can be defined by the following expansion:
\be
\wp(z|\tau)=\frac{1}{z^2}+\sum_{m,n\neq(0,0)}\biggl(\frac{1}{(z+m+n\tau)^2}-
\frac{1}{(m+n\tau)^2}\biggr).
 \ee
This is a periodic function with periods $\tau$ and $1$ and with a
double pole at $z=0$. This function solves the following
differential equation: \be
\wp'(z|\tau)^2=4(\wp(z|\tau)-e_1)(\wp(z|\tau)-e_2)(\wp(z|\tau)-e_3),\quad
e_1+e_2+e_3=0\ee where the derivative is with respect to $z$,
$e_i\equiv \wp(\omega_i)$ and $\omega_i$ are the half-periods of
the torus, \be \omega_1=\frac{1}{2},\quad
\omega_2=\frac{1+\tau}{2},\quad \omega_3=\frac{\tau}{2}. \ee

Another useful property of this function is the addition theorem:
\be
\wp(z_1+z_2)=\frac{1}{4}\biggl(\frac{\wp'(z_1)-\wp'(z_2)}{\wp(z_1)-\wp(z_2)}\biggr)^2-\wp(z_1)-\wp(z_2).
\ee

Any elliptic function  $f(z|\tau)$ can be written as:
\be f(z|\tau)=P(\wp(z|\tau))+\wp'(z|\tau)Q(\wp(z|\tau))\ee
where $P(x)$ and $Q(x)$ are rational functions. From the properties
we quoted above it follows that if one is given the
structure of singularities of a given elliptic function, and
one succeeds in building another elliptic function with the same
singularities, the two functions will differ by a constant. We
will use this fact extensively.

There are some additional functions related to the Weierstrass
function, which are not elliptic but nevertheless play an
important role. We define $\sigma(z)$ and $\zeta(z)$ by their
relation to the Weierstrass $\wp(z)$ function, \be
\frac{\s'(z)}{\s(z)}=\zeta(z),\quad\quad \zeta'(z)=-\wp(z). \ee A
useful identity for $\s(z)$ is its variation along the periods of
the torus
\begin{gather}
\s(z+2\omega_i)=-e^{2\eta_i(z+\omega_i)}\s(z),
\end{gather}
where $\eta_i \equiv \zeta(\omega_i)$.
We will also use the Legendre theorem, which states that
for the periods $(2\omega_1,2\omega_3)=(1,\tau)$
\be
\tau\eta_1-\eta_3=i\pi.
\ee

\bibliographystyle{amsplain}

\begin{thebibliography}{10}
\bibitem{'tHooft:1973jz}
G.~'t Hooft, ``A planar diagram theory for strong interactions,''
Nucl.\ Phys.\ B {\bf 72}, 461 (1974).
\bibitem{Maldacena:1997re}
  J.~M.~Maldacena,
  ``The large N limit of superconformal field theories and supergravity,''
  Adv.\ Theor.\ Math.\ Phys.\  {\bf 2}, 231 (1998)
  [Int.\ J.\ Theor.\ Phys.\  {\bf 38}, 1113 (1999)]
  [arXiv:hep-th/9711200];
%
  S.~S.~Gubser, I.~R.~Klebanov and A.~M.~Polyakov,
  ``Gauge theory correlators from non-critical string theory,''
  Phys.\ Lett.\ B {\bf 428} (1998) 105
  [arXiv:hep-th/9802109];
%
  E.~Witten,
  ``Anti-de Sitter space and holography,''
  Adv.\ Theor.\ Math.\ Phys.\  {\bf 2}, 253 (1998)
  [arXiv:hep-th/9802150].
\bibitem{GopakumarVafa}
  R.~Gopakumar and C.~Vafa,
  ``On the gauge theory/geometry correspondence,''
  Adv.\ Theor.\ Math.\ Phys.\  {\bf 3}, 1415 (1999)
  [arXiv:hep-th/9811131];
  H.~Ooguri and C.~Vafa,
  ``Worldsheet derivation of a large N duality,''
  Nucl.\ Phys.\ B {\bf 641}, 3 (2002)
  [arXiv:hep-th/0205297].
\bibitem{ConfDeconf}
  B.~Sundborg,
  ``The Hagedorn transition, deconfinement and N = 4 SYM theory,''
  Nucl.\ Phys.\ B {\bf 573}, 349 (2000)
  [arXiv:hep-th/9908001];
  O.~Aharony, J.~Marsano, S.~Minwalla, K.~Papadodimas and M.~Van Raamsdonk,
  ``The Hagedorn / deconfinement phase transition in weakly coupled large N
  gauge theories,''
  Adv.\ Theor.\ Math.\ Phys.\  {\bf 8}, 603 (2004)
  [arXiv:hep-th/0310285].
\bibitem{FreeFields}
K.~Bardakci and C.~B.~Thorn,
``A worldsheet description of large $N_c$ quantum field theory,''
Nucl.\ Phys.\ B {\bf 626}, 287 (2002)
[arXiv:hep-th/0110301]; C.~B.~Thorn,
``A worldsheet description of planar Yang-Mills theory,''
Nucl.\ Phys.\ B {\bf 637}, 272 (2002)
[Erratum-ibid.\ B {\bf 648}, 457 (2003)]
[arXiv:hep-th/0203167]; K.~Bardakci and C.~B.~Thorn,
``A mean field approximation to the world sheet model of planar $\phi^3$ field
theory,''
Nucl.\ Phys.\ B {\bf 652}, 196 (2003)
[arXiv:hep-th/0206205]; S.~Gudmundsson, C.~B.~Thorn and T.~A.~Tran,
``BT worldsheet for supersymmetric gauge theories,''
Nucl.\ Phys.\ B {\bf 649}, 3 (2003)
[arXiv:hep-th/0209102]; K.~Bardakci and C.~B.~Thorn,
``An improved mean field approximation on the worldsheet for planar $\phi^3$
theory,''
Nucl.\ Phys.\ B {\bf 661}, 235 (2003)
[arXiv:hep-th/0212254]; C.~B.~Thorn and T.~A.~Tran,
``The fishnet as anti-ferromagnetic phase of worldsheet Ising spins,''
Nucl.\ Phys.\ B {\bf 677}, 289 (2004)
[arXiv:hep-th/0307203];
  K.~Bardakci,
  ``Further results about field theory on the world sheet and string
  formation,''
  Nucl.\ Phys.\ B {\bf 715} (2005) 141
  [arXiv:hep-th/0501107].
\bibitem{StringBits}
  P.~Haggi-Mani and B.~Sundborg,
  ``Free large $N$ supersymmetric Yang-Mills theory as a string theory,''
  JHEP {\bf 0004} (2000) 031 [arXiv:hep-th/0002189];
  H.~L.~Verlinde,
  ``Bits, matrices and $1/N$,''
  JHEP {\bf 0312} (2003) 052
  [arXiv:hep-th/0206059];
  J.~G.~Zhou,
  ``pp-wave string interactions from string bit model,''
  Phys.\ Rev.\ D {\bf 67} (2003) 026010
  [arXiv:hep-th/0208232];
  D.~Vaman and H.~L.~Verlinde,
  ``Bit strings from N = 4 gauge theory,''
  JHEP {\bf 0311} (2003) 041
  [arXiv:hep-th/0209215];
  A.~Dhar, G.~Mandal and S.~R.~Wadia,
  ``String bits in small radius AdS and weakly coupled N = 4 super  Yang-Mills
  theory. I,''
  arXiv:hep-th/0304062;
  K.~Okuyama and L.~S.~Tseng,
  ``Three-point functions in N = 4 SYM theory at one-loop,''
  JHEP {\bf 0408} (2004) 055 [arXiv:hep-th/0404190];
  L.~F.~Alday, J.~R.~David, E.~Gava and K.~S.~Narain,
  ``Structure constants of planar N = 4 Yang Mills at one loop,''
  JHEP {\bf 0509} (2005) 070
  [arXiv:hep-th/0502186];
  J.~Engquist and P.~Sundell,
  ``Brane partons and singleton strings,''
  arXiv:hep-th/0508124;
  L.~F.~Alday, J.~R.~David, E.~Gava and K.~S.~Narain,
  ``Towards a string bit formulation of N = 4 super Yang-Mills,''
  arXiv:hep-th/0510264.

\bibitem{Joe}
J.~Polchinski, unpublished.

\bibitem{Karch}
A.~Karch,
``Lightcone quantization of string theory duals of free field
theories,'' arXiv:hep-th/0212041;
A.~Clark, A.~Karch, P.~Kovtun and D.~Yamada,
``Construction of bosonic string theory on infinitely curved
anti-de  Sitter space,'' Phys.\ Rev.\ D {\bf 68}, 066011 (2003)
[arXiv:hep-th/0304107].

\bibitem{Bonelli:2004ve}
G.~Bonelli, ``On the boundary gauge dual of closed tensionless
free strings in AdS,'' JHEP {\bf 0411} (2004) 059
[arXiv:hep-th/0407144].

\bibitem{Itzhaki:2004te}
  N.~Itzhaki and J.~McGreevy,
  ``The large N harmonic oscillator as a string theory,''
  Phys.\ Rev.\ D {\bf 71} (2005) 025003
  [arXiv:hep-th/0408180].
\bibitem{Gopakumars}
R.~Gopakumar, ``From free fields to AdS,'' Phys.\ Rev.\ D {\bf
70}, 025009 (2004) [arXiv:hep-th/0308184];
R.~Gopakumar, ``From free fields to AdS. II,'' Phys.\ Rev.\ D {\bf
70}, 025010 (2004) [arXiv:hep-th/0402063];
R.~Gopakumar, ``Free field theory as a string theory?,'' Comptes
Rendus Physique {\bf 5}, 1111 (2004) [arXiv:hep-th/0409233];
R.~Gopakumar, ``From free fields to AdS. III,'' Phys.\ Rev.\ D
{\bf 72}, 066008 (2005) [arXiv:hep-th/0504229].
\bibitem{Bianchi}
  M.~Bianchi, J.~F.~Morales and H.~Samtleben,
  ``On stringy AdS(5) x S**5 and higher spin holography,''
  JHEP {\bf 0307} (2003) 062
  [arXiv:hep-th/0305052];
  N.~Beisert, M.~Bianchi, J.~F.~Morales and H.~Samtleben,
  ``On the spectrum of AdS/CFT beyond supergravity,''
  JHEP {\bf 0402} (2004) 001
  [arXiv:hep-th/0310292].
\bibitem{Akhmedov}
  E.~T.~Akhmedov,
  ``Expansion in Feynman graphs as simplicial string theory,''
  JETP Lett.\  {\bf 80}, 218 (2004)
  [Pisma Zh.\ Eksp.\ Teor.\ Fiz.\  {\bf 80}, 247 (2004)]
  [arXiv:hep-th/0407018].
\bibitem{Furuuchi:2005qm}
  K.~Furuuchi,
  ``From free fields to AdS: Thermal case,''
  Phys.\ Rev.\ D {\bf 72} (2005) 066009
  [arXiv:hep-th/0505148].
\bibitem{Zwiebach}
B.~Zwiebach,
  ``Closed string field theory: Quantum action and the B-V master equation,''
  Nucl.\ Phys.\ B {\bf 390}, 33 (1993)
  [arXiv:hep-th/9206084];
A.~Belopolsky and B.~Zwiebach,
  ``Off-shell closed string amplitudes: Towards a computation of the tachyon
  potential,''
  Nucl.\ Phys.\ B {\bf 442}, 494 (1995)
  [arXiv:hep-th/9409015].
\bibitem{Moeller:2004yy}
N.~Moeller, ``Closed bosonic string field theory at quartic
order,'' JHEP {\bf 0411} (2004) 018 [arXiv:hep-th/0408067].
\bibitem{K.Strebel:1984}
K.~Strebel, ``Quadratic differentials,'' Springer-Verlag, 1984.
\bibitem{Mulase:98}
M.~Mulase, M.~Penkava, ``Ribbon graphs, quadratic differentials on
Riemann surfaces, and algebraic curves defined over $\bar Q$ ,''
[math-ph/9811024].
\bibitem{Zvonkine:2002}
D.~Zvonkine, ``Strebel differentials on stable curves and
Kontsevich's proof of Witten's conjecture,'' [math.AG/0209071].
\bibitem{Konts}
  M.~Kontsevich,
  ``Intersection theory on the moduli space of curves and the matrix Airy
  function,''
  Commun.\ Math.\ Phys.\  {\bf 147}, 1 (1992);
  E.~Witten,
  ``Two-dimensional gravity and intersection theory on moduli space,''
  Surveys Diff.\ Geom.\  {\bf 1}, 243 (1991).
\bibitem{Teschner}
  J.~Teschner,
  ``On structure constants and fusion rules in the SL(2,C)/SU(2) WZNW  model,''
  Nucl.\ Phys.\ B {\bf 546}, 390 (1999)
  [arXiv:hep-th/9712256];
  J.~Teschner,
  ``The mini-superspace limit of the SL(2,C)/SU(2) WZNW model,''
  Nucl.\ Phys.\ B {\bf 546}, 369 (1999)
  [arXiv:hep-th/9712258];
  J.~Teschner,
  ``Operator product expansion and factorization in the $H_3^+$ WZNW model,''
  Nucl.\ Phys.\ B {\bf 571}, 555 (2000)
  [arXiv:hep-th/9906215];
   J.~Teschner,
  ``Crossing symmetry in the $H_3^+$ WZNW model,''
  Phys.\ Lett.\ B {\bf 521}, 127 (2001)
  [arXiv:hep-th/0108121].
\bibitem{MaldacenaOoguri}
  J.~M.~Maldacena and H.~Ooguri,
  ``Strings in AdS(3) and SL(2,R) WZW model. I,''
  J.\ Math.\ Phys.\  {\bf 42}, 2929 (2001)
  [arXiv:hep-th/0001053];
  J.~M.~Maldacena, H.~Ooguri and J.~Son,
  ``Strings in AdS(3) and the SL(2,R) WZW model. II: Euclidean black hole,''
  J.\ Math.\ Phys.\  {\bf 42}, 2961 (2001)
  [arXiv:hep-th/0005183];
  J.~M.~Maldacena and H.~Ooguri,
  ``Strings in AdS(3) and the SL(2,R) WZW model. III: Correlation  functions,''
  Phys.\ Rev.\ D {\bf 65}, 106006 (2002)
  [arXiv:hep-th/0111180].
\bibitem{GiveonKutasov}
  A.~Giveon, D.~Kutasov and O.~Pelc,
  ``Holography for non-critical superstrings,''
  JHEP {\bf 9910} (1999) 035
  [arXiv:hep-th/9907178];
  A.~Giveon and D.~Kutasov,
  ``Little string theory in a double scaling limit,''
  JHEP {\bf 9910} (1999) 034
  [arXiv:hep-th/9909110];
  A.~Giveon and D.~Kutasov,
  ``Comments on double scaled little string theory,''
  JHEP {\bf 0001} (2000) 023
  [arXiv:hep-th/9911039].
\bibitem{Kutasov:1999xu}
  D.~Kutasov and N.~Seiberg,
  ``More comments on string theory on AdS(3),''
  JHEP {\bf 9904} (1999) 008
  [arXiv:hep-th/9903219].
\bibitem{Constable:2002hw}
  N.~R.~Constable, D.~Z.~Freedman, M.~Headrick, S.~Minwalla, L.~Motl,
A.~Postnikov and W.~Skiba,
  ``PP-wave string interactions from perturbative Yang-Mills theory,''
  JHEP {\bf 0207} (2002) 017
  [arXiv:hep-th/0205089].
\end{thebibliography}

\end{document}